\documentclass[sigconf, nonacm]{acmart}

\usepackage[
textsize=footnotesize]{todonotes}
\presetkeys{todonotes}{inline}{}
\usepackage{algorithmic}
\usepackage{graphicx}
\usepackage{textcomp}
\usepackage{xcolor}
\usepackage{booktabs}
\usepackage{hyperref}
\usepackage{paralist}
\usepackage{tabularx}
\usepackage{ragged2e}  % for '\RaggedRight' macro (allows hyphenation)
\usepackage{lscape}
\usepackage{listings}
\usepackage{balance} % To balance out last page

\newcommand{\Darwin}{\emph{Darwin}}
\newcommand{\MigCast}{\emph{MigCast}}
\newcommand{\MCMC}{\emph{MigCast in Monte Carlo}}

\newcommand{\nop}[1]{{}} 
\newtheorem{observation}{Observation}

\begin{document}

\title[\emph{MigCast in Monte Carlo}]{\emph{MigCast in Monte Carlo}:\\The Impact of Data Model Evolution\\in NoSQL Databases}

\author{Andrea Hillenbrand}
\affiliation{%
    \institution{Darmstadt University of Applied Sciences}
    \city{Darmstadt}
    \country{Germany}
    }
\email{andrea.hillenbrand@h-da.de}

\author{Uta St\"orl}
\affiliation{%
    \institution{University of Hagen}
    \city{Hagen}
    \country{Germany}
    }
\email{uta.stoerl@fernuni-hagen.de}

\author{Shamil Nabiyev}
\affiliation{%
    \institution{Darmstadt University of Applied Sciences}
    \city{Darmstadt}
    \country{Germany}
    }
\email{bdcc@h-da.de}

\author{Stefanie Scherzinger}
\affiliation{%
    \institution{University of Passau}
    \city{Passau}
    \country{Germany}
    }
\email{stefanie.scherzinger@uni-passau.de}

\begin{abstract}
During the development of NoSQL-backed software, the data model evolves naturally alongside the application code. Especially in agile development, new application releases are deployed frequently causing schema changes. Eventually, decisions have to be made regarding the migration of versioned legacy data which is persisted in the cloud-hosted production database. We solve this schema evolution problem and present the results of near-exhaustive calculations by means of which software project stakeholders can manage the operative costs for data model evolution and adapt their software release strategy accordingly in order to comply with service-level agreements regarding the competing metrics of migration costs and latency. We clarify conclusively how data model evolution in NoSQL databases impacts the metrics while taking all relevant characteristics of migration scenarios into account. As calculating all possible combinatorics in the search space of migration scenarios would by far exceed computational means, we used a probabilistic Monte Carlo method of repeated sampling, serving as a well-established means to bring the complexity of data model evolution under control. Our experiments show the qualitative and quantitative impact on the performance of migration strategies with respect to intensity and distribution of data entity accesses, the kinds of schema changes, and the characteristics of the underlying data model.
\end{abstract}

\maketitle

%\keywords{Databases; NoSQL; Data Migration; Schema Evolution; Risk Management; Monte Carlo Simulation; Migration Cost; Latency} % 

%%%%%%%%%%%%%%%%%%%%%%%%%%%%%%%%%%%%%%%%%%%%%%%%%%%%%%%
%%%%%%%%%%%%%%% SECTION 1: INTRODUCTION %%%%%%%%%%%%%%% 
%%%%%%%%%%%%%%%%%%%%%%%%%%%%%%%%%%%%%%%%%%%%%%%%%%%%%%%

\section{Introduction}
\label{sec:introduction}

Developing or maintaining a software-as-a-service application requires the management of steadily increasing amounts of data and their co-evolution with the software code~\cite{Qiu2013,Goeminne2014}. In this context, schema-flexible NoSQL databases have become especially popular backends in agile development. NoSQL databases allow application developers to write code assuming a new data model that is different from the current database schema~\cite{Scherzinger2013,Davoudian2018}. Furthermore, new software releases can be deployed without migration-related application downtime. In fact, a very recent empirical study on the evolution of NoSQL database schemas has shown that software releases include considerably more schema-relevant changes (>30\% compared to 2\% with relational databases)~\cite{Scherzinger2020}. Furthermore, NoSQL database schemas generally grow in complexity over time just like relational database schemas, however, they take longer to stabilize. Arguably, it can be deduced that schemas in NoSQL databases \emph{evolve} more flexibly alongside the software application code.

Though, eventually, the issue of handling the variational data in the NoSQL database has to be addressed. Imagine as a software project stakeholder, you are planning the next software release and need to take all data into account that is persisted in your cloud-hosted production database. A decision has to be made as to when to migrate which legacy data that is structured according to earlier schema versions. %Regardless of whether your NoSQL database calls itself schema-less, the new application code assumes a different data model in respect of what is being stored in the production database. 
If all of the legacy data is curated according to the latest data model in one go at the release of schema changes, i.e., with an \emph{eager} migration strategy, then maximal charges are produced with the cloud provider. In addition, a lower application performance must be expected during the migration process in case that application downtime needs to be avoided. 

Currently, cloud providers such as \emph{Google Cloud} charge for database reads, writes, and deletes, which we have realized in our cost model. Note that simply adding a field to the data schema requires updating and rewriting of all affected data entities in the database. Curating all variational data by eagerly migrating all legacy entities can significantly drive up the operational costs for the cloud service, especially in case of frequent software releases. For instance, for 10M persisted entities on \emph{Google Cloud} and 100 monthly schema changes on average throughout a year, charges of approx. USD 13,000 are incurred just for database writes alone\footnote{Writing an entity currently costs USD 0.108 per 100,000 documents for regional location pricing in North America as of March 10, 2021 (see \url{https://cloud.google.com/datastore/pricing}). Not all schema changes add properties to the entities, yet we assume this rough estimate here, because there are both cheaper schema changes (deletes) as well as more expensive schema changes like reorganizing properties which affect multiple types and thus run up higher charges. As to an empirical analysis on the changes over time for relational databases refer to~\cite{Qiu2013}, and for NoSQL databases to~\cite{Scherzinger2020}.}---not even counting costs for database reads or for storing the data.

The benefit of this investment is that the application then accesses a structurally homogeneous database instance. Then, reads and writes against the database come at no migration-induced overhead accounting for structural variety and hence, the time that it takes for the requested data entities to be retrieved, the \emph{data access latency}, is minimal. Short access times are indeed crucial to application performance, especially in cloud-hosted applications~\cite{Curino2011,Chen2016,Difallah2013,Barker2012}.

\begin{figure}[ht]
\centering
    \includegraphics[trim=10 9 8 8,clip,width=0.475\textwidth]{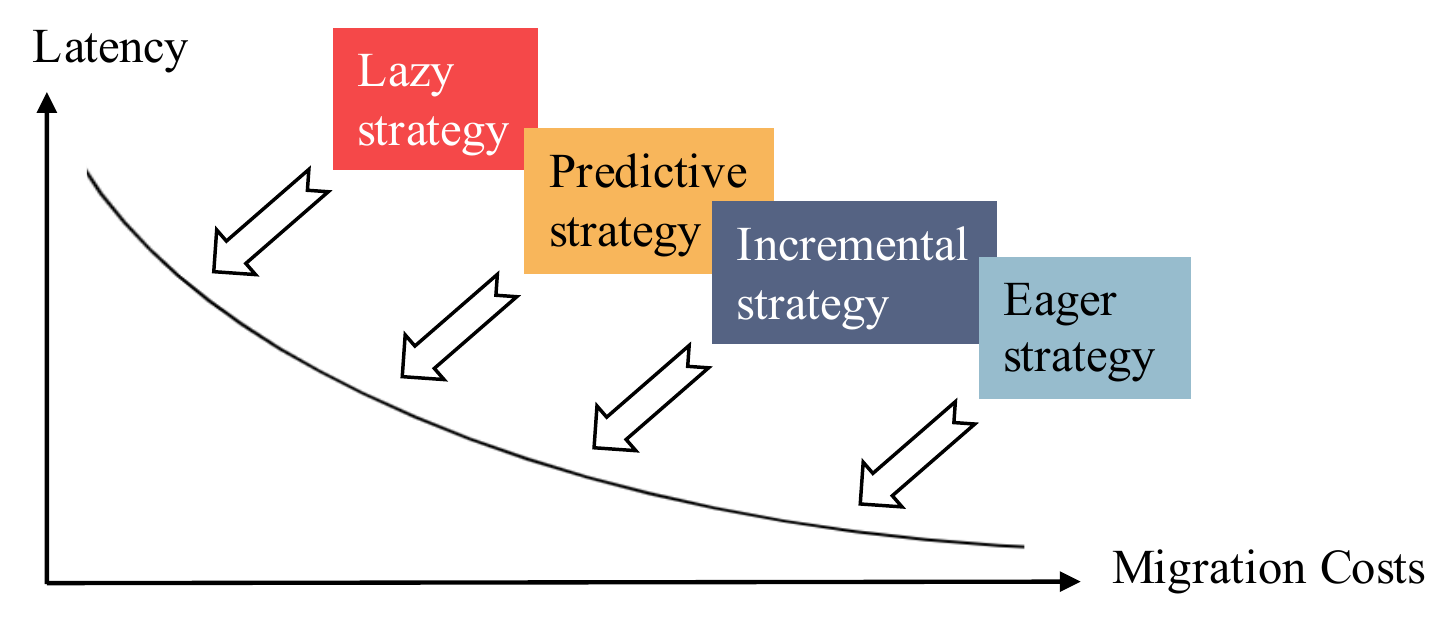}
	\caption{Tradeoff in choosing a  migration strategy (schema-tic representation of Pareto-distributed data accesses).}%~\cite{Hillenbrand2019}.}
	\label{fig:tradeoff}
\end{figure}

If latency is the sole criterion, then an \emph{eager} migration strategy is clearly most suitable. If, on the other hand, saving costs is the most important criterion, then a \emph{lazy} migration strategy should be applied which minimizes \emph{migration costs}, as data remains unchanged in the event of a release. In this case, if a legacy entity is accessed, it is migrated individually and on-the-fly, then being in congruence with the latest schema version, yet introducing a considerable runtime overhead~\cite{scdm16,Hillenbrand2019,Saur2016}. These two metrics, \emph{migration costs} and \emph{data access latency}, are in fact competitors in a tradeoff, which is schematically depicted in Figure~\ref{fig:tradeoff}. The metrics cannot be optimized independently from each other and thus, a choice of a migration strategy is also a choice on the tradeoff between these metrics at different opportunity costs for alternative migration strategies.

In fact, the growing importance of not only cost but also energy efficiency is reflected in the dedication of entire workshops to \emph{Green Data Centers}~\cite{DataCenters,DataCenter2}. Similarly, service-level agreements are oftentimes formulated as \emph{Green SLAs}~\cite{GreenSLA}, that is, as SLAs put in place in order to save energy. Although the correlation between migration costs and energy consumption is still being researched, it can surely be assumed that a reduction of data migration (and thus of migration costs) implies a reduction of energy consumption (and thus of the $CO_2$ footprint caused by this migration). Thus, the tradeoff between the metrics of migration costs and latency should be decided upon while keeping in mind that any compromise also affects the application system's energy consumption. In any case, a decision by a software project stakeholder on the migration of legacy entities can only then be realized as a certain cost-aware compromise that complies with latency-related SLAs, if the relationship between these metrics is clarified and respective opportunity costs are completely transparent, which we contribute in this paper.

In fact, the data management research community has identified data model evolution as one of the hardest problems that is currently being ignored~\cite{Stonebraker2018}. Especially when a database is considered to be run in a cloud-hosted environment, concerns about high costs are most often cited in surveys, while increased latency being an issue as well~\cite{MongoDB2020}. We argue that the lack of cost transparency regarding data migration costs and latency should be a crucial issue to all stakeholders. In this paper, we conclusively investigate and clarify the impact of schema evolution on both metrics for common migration strategies in respect of all relevant migration scenarios. %\emph{eager}, \emph{lazy}, \emph{incremental}, and \emph{predictive} migration strategies (refer to Subsection~\ref{subsec:evolution} for a definition). 
%Thereby, we equip software project stakeholders to make informed decisions when choosing a migration strategy adequate to the migration scenario and in compliance with service-level agreements. Furthermore, we discuss the options that stakeholders have to influence these metrics further by means of adapting the software release strategy in order to remain in control of the consequences of a flexibly evolving data model.

Yet, solving the schema evolution problem is not trivial, because data migration scenarios are highly complex as they rely on many factors that influence the impact on the metrics. Completely searching the solution space would by far exceed computational means. We use a probabilistic \emph{Monte Carlo} method of repeated sampling to investigate the scenario characteristics while bringing the complexity of the problem under control~\cite{Mackay2003,Fishman2013} We parameterize each scenario and calculate the metrics for each migration strategy by means of the tool-based migration advisor \MigCast{}, which we presented in earlier work~\cite{Hillenbrand2019,Hillenbrand2020}.  Our \MigCast{} tool stands on the tradition of advisors in database technology~\cite{Chaudhuri2009} by maintaining a cost model and forecasting migration costs for alternative data migration strategies. By repeatedly sampling all relevant migration scenarios and randomizing some algorithm parameters---the Monte Carlo method---we uncover the correlations of the scenario characteristics and their impact on the metrics. Probabilistic approaches have been applied successfully regarding research questions in database management~\cite{Aggarwal2009,Haas2018,Jampani2008,Jampani2011,Sharma2019,Suciu2011,Dalvi2007}.

\paragraph{Contributions.}\label{contributions}
Our paper makes the following contributions:
\begin{itemize}
    \item By means of near-exhaustive calculations, we investigate in depth and clarify conclusively how schema evolution in NoSQL databases impacts the competing metrics migration costs and latency. We investigated all relevant migration scenario characteristics with an underlying cost model that takes into account (i.) intensity and distribution of data entity accesses, (ii.) different kinds of schema changes, and (iii.)  characteristics of the data model.
    \item We present the results and discuss the implications such that software project stakeholders can manage and control the operative costs for schema evolution and adapt the pace of their software release strategy accordingly in order to ascertain the compliance with cost- and latency-related service-level agreements.
    \item We are the first to apply a probabilistic Monte Carlo method of repeated sampling in order to bring the inherent complexity of the schema evolution problem under control.
    %\item Our experiments show the qualitative and quantitative impact of schema evolution on the metrics and thus characterize the performance of migration strategies with respect to all relevant migration scenario characteristics.
    \item For all relevant scenario characteristics, we (i.) quantify the average opportunity costs for each migration strategy, (ii.) identify multi-type schema modification operations as the cost driver of schema evolution, (iii.) find a predictive migration strategy utilizing the Pareto principle to be the best compromise between the metrics with regard to SLA compliance, and (iv.) identify high cardinality of the relationships of the underlying data model as the cause for a high variance of the impacts on the metrics.
%    \item We clarify definitively what the characteristics of the most popular migration strategies are depending on the particular migration scenario.
    %\item persistence of the simulation results to ensure their reproducibility, understandability, and comparability, thus contributing to data provenance;
    %\item and we deliver a detailed evaluation of passed or failed simulation invariants ensuring model consistency and simulation validity.
\end{itemize}

\paragraph{Structure.} This present paper is structured as follows: In the preliminaries, Section~\ref{sec:preliminaries}, we clarify the schema evolution problem, introduce the migration strategies and our terminology, and motivate the Monte Carlo approach. Afterwards, we address related work in Section~\ref{sec:related}. In Section~\ref{sec:architecture}, the system architecture and methodology of the experiment settings are specified. A detailed presentation of the experiment results follows in Section~\ref{sec:results}. In Section~\ref{sec:discussion}, we evaluate the results and discuss their implications, before we conclude this paper in Section~\ref{sec:conclusion}.

%%%%%%%%%%%%%%%%%%%%%%%%%%%%%%%%%%%%%%%%%%%%%%%%%%%%%%%
%%%%%%%%%%%%%%% SECTION 2: PRELIMINARIES %%%%%%%%%%%%%% 
%%%%%%%%%%%%%%%%%%%%%%%%%%%%%%%%%%%%%%%%%%%%%%%%%%%%%%%

\section{Preliminaries}
\label{sec:preliminaries}

In this section, we clarify the schema evolution problem, introduce the investigated data migration strategies as well as the used terminology, and motivate the Monte Carlo approach.

\subsection{The Schema Evolution Problem}
\label{subsec:scenario}

\begin{figure*}[ht]
\includegraphics[trim=5 6 4 8,clip,width=\textwidth]{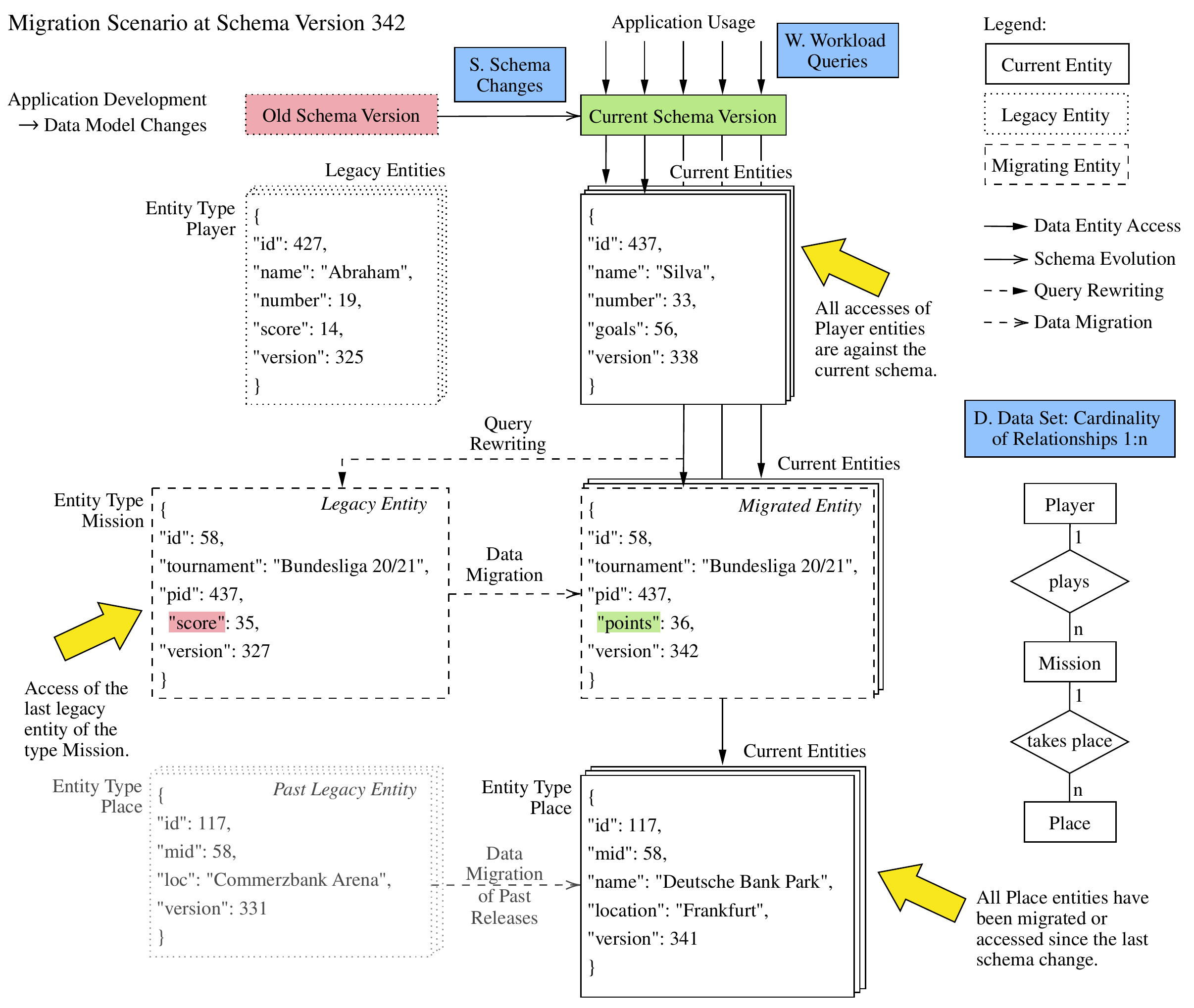}
\caption{Migration scenario after a schema modification operation had renamed a property of a type; highlighted in blue are the investigated migration scenario characteristics (further details in Sections~\ref{sec:architecture} and~\ref{sec:results}).}
\label{fig:scenario}
\end{figure*}

Figure~\ref{fig:scenario} illustrates the data model of an online game that we used in our \MCMC{} experiments. The data model of the application has the entity types \emph{Player}, \emph{Mission}, and \emph{Place} (see D.). In the figure, a migration scenario is shown at a certain schema version. The application code queries data entities against the current schema version. During the development of the application, the data model has undergone many changes (see S.) so that variational data, which still abides by older schema versions, has to be curated when legacy entities need to be accessed (see W.). These queries then have to be rewritten and are migrated on-the-fly in order to match the current schema version. In the figure, a legacy entity of the type \emph{Mission} is currently being updated, whose schema was subject of a modification operation that renamed the property \emph{score} to \emph{points}. Thus, the entity is migrated to match the current schema, which causes \emph{on-read migration costs}. When data entities become legacy entities through a schema change and are instantaneously migrated to match this schema version, then we count these migration costs towards \emph{on-release costs}.

In order to investigate the impact of schema evolution on the competing metrics of migration costs and latency, we base the calculations of \MigCast{} on a cost model that takes into account (W.) the intensity and distribution of data entity accesses, (S.) the different kinds of schema changes, and (D.) the cardinality of the relationships of the underlying data model. These investigated characteristics are highlighted in Figure~\ref{fig:scenario} in blue and are varied as detailed in Table~\ref{table:varied} on page~\pageref{table:varied}.

\subsection{Investigated Data Migration Strategies}
\label{subsec:evolution}

A migration strategy has to be decided for when schema changes are implied through new releases of software application code. Each migration strategy handles the migration of legacy data differently, thereby settling on a different compromise regarding the tradeoff between the competing metrics of migration costs and latency. We define \emph{migration costs} as the charges occasioned by migrating the legacy data according to the most current data model and \emph{(data access) latency} as the time that it takes for the data to be retrieved.

Through the \emph{eager} migration strategy, all legacy entities are migrated according to the latest schema version. Then, latency is optimal as the application code can access a structurally homogeneous database instance, yet migration costs are maximal. With \emph{lazy} migration, legacy data remains unchanged in the event of a release, incurring minimal migration charges, yet introducing a runtime overhead on reads and writes~\cite{Hillenbrand2019}. The migration strategies are complemented by two more \emph{proactive} strategies, which act in advance of situations when migrating legacy entities could cause significant latency overhead, the \emph{incremental} and the \emph{predictive} migration strategy. With these strategies, a compromise is reached between the competing metrics.

The \emph{incremental} migration strategy migrates all legacy entities at certain points in time. Lazy periods of time are then interrupted by regular bouts of tidying up the structurally heterogeneous database instance in order to get rid of the runtime overhead intermittently. In order to keep a steady balance on the tradeoff between migration costs and latency, \emph{predictive} migration is applied when new application code is released that includes schema changes. \emph{Predictive} migration allows improving latency at moderate costs in case that data entity accesses are \emph{Pareto} distributed, i.e., concentrate on particular entities. Then, the prediction is based on the assumption that the oftener entities were accessed in the past, the more likely it is that they be accessed again in the future. The \emph{predictive} migration strategy is implemented by keeping track of past data accesses and ordering the accessed entities accordingly via \emph{exponential smoothing}. This established technique in time series data weighs the entities by their actuality and access frequency~\cite{DBLP:conf/icde/LevandoskiLS13}: The more recent the entity accesses, the higher the weight of the entity. Weights decrease exponentially over time simulating an aging process of the entities by accounting for actuality as well as for access frequency. %, thus reaching a subtler compromise compared to the previously discussed migration strategies. %The ordered, to-be-migrated entities are kept in the \emph{prediction set}. By this approach a subtle compromise is being reached on the tradeoff between migration costs and latency compared to the previously discussed migration strategies. 

%As a final metric, the \emph{Migration Debt} is also included in our observations, yet not discussed extensively in this present paper. It refers to the charges that would have to be invested in order to migrate all legacy entities to a structurally homogeneous database instance, based on the past on-release and on-read migration costs and the number of remaining legacy entities.

%We refer to the data entity accesses in between two releases of schema changes as \emph{workload}.

\subsection{Motivating a Monte Carlo Approach}
\label{subsec:montecarlo}

In order to enable software project stakeholders to manage the impact of schema evolution, it has to be investigated how relevant factors influence the metrics migration costs and latency. A straightforward approach would be to assume averages for each of the parameters describing a migration scenario, but that would not provide information how the metrics vary and potentially obscure a correlation between scenario characteristics and their impact on the metrics. Furthermore, averages are not reasonably applicable in case of data entity accesses. However, by means of the \emph{Monte Carlo} approach~\cite{Upton2002,Mackay2003,Fishman2013}, that is, an approach utilizing the \emph{Monte Carlo method}, this correlation can be uncovered. It is a well-established approach used in the data management context~\cite{Aggarwal2009,Haas2018, Jampani2008,Jampani2011,Sharma2019,Suciu2011,Dalvi2007}. 

The Monte Carlo method is used to solve problems that are deterministic as such, but more efficiently solved by probabilistic means, that is, when there are too many possibilities to be calculated exceeding the computational means, when relevant input variables of the computations are unknown or their acquisition costly~\cite{Upton2002,Mackay2003,Fishman2013}. Then, a repeated sampling of the input variables of the deterministic algorithm uses random values during the calculation in order to obtain a distribution of results (refer to Subsection~\ref{subsec:architecture} as to how input parameters are randomized). The \emph{sample mean} of a metric can then be interpreted as the most probable prediction~\cite{Upton2002,Mackay2003,Fishman2013}. By knowing this sample mean and the possible range of the metrics, stakeholders are enabled to assess and effectively manage the potential risk that usually comes along with a lack of cost transparency and possible variance of the metrics. When applying the Monte Carlo method to our use case of schema evolution, we had to decide how often the sampling needs to be repeated so that the results can be generalized. We now introduce the necessary notions in this context and then return to this question.

\begin{figure}[ht]
\raggedright
\includegraphics[trim=8 8 10 3,clip,width=0.475\textwidth]{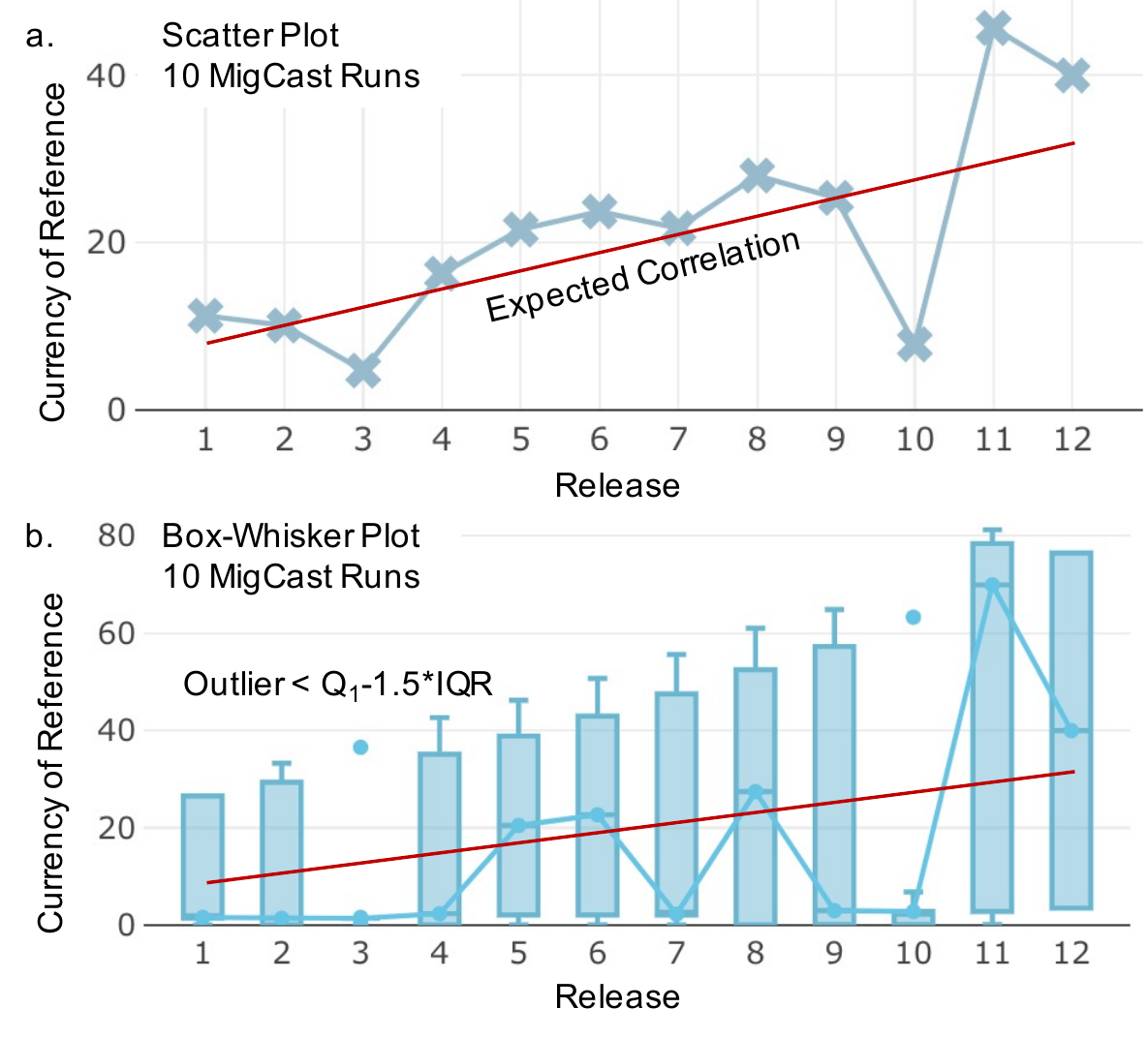}
\caption{Projected on-release migration costs of 12 releases of schema changes for the eager migration strategy.}%; Configuration: Pareto-distributed, medium workload, 25\% multi-type SMO complexity, high cardinality.}
\label{fig:regression}
\end{figure}

In Figure~\ref{fig:regression}, the projected migration costs of 12 releases of schema changes are depicted in a Scatter plot (a.) and a Box-Whisker plot (b.) for the \emph{eager} migration strategy. The projections are based on 10 runs of a Monte Carlo approach using the migration advisor \MigCast{}. Release number 11 shows a rather high \emph{sample mean}. Sample means in the Scatter plot are implemented as \emph{arithmetic averages} and in the Box-Whisker plot as \emph{medians}. Outliers are depicted in our Box-Whisker plot as dots and taken into account when calculating the median. If there are no outliers depicted, then all calculated values are within a certain range indicated as boxes, the \emph{interquartile range} (IQR) or \emph{mid-range}~\cite{Upton2002}, together with their whiskers. The whiskers determine a range calculated through subtracting the absolute value of the IQR, multiplied by a factor 1.5, from the lower quartile $Q_1$ and adding the very same value to the upper quartile $Q_3$, i.e., $Q_1-1.5\ |\operatorname{IQR}$| and $Q_3+1.5\ |\operatorname{IQR}|$.

\begin{figure}[ht]
\raggedright
\includegraphics[trim=4 6 12 6,clip,width=0.475\textwidth]{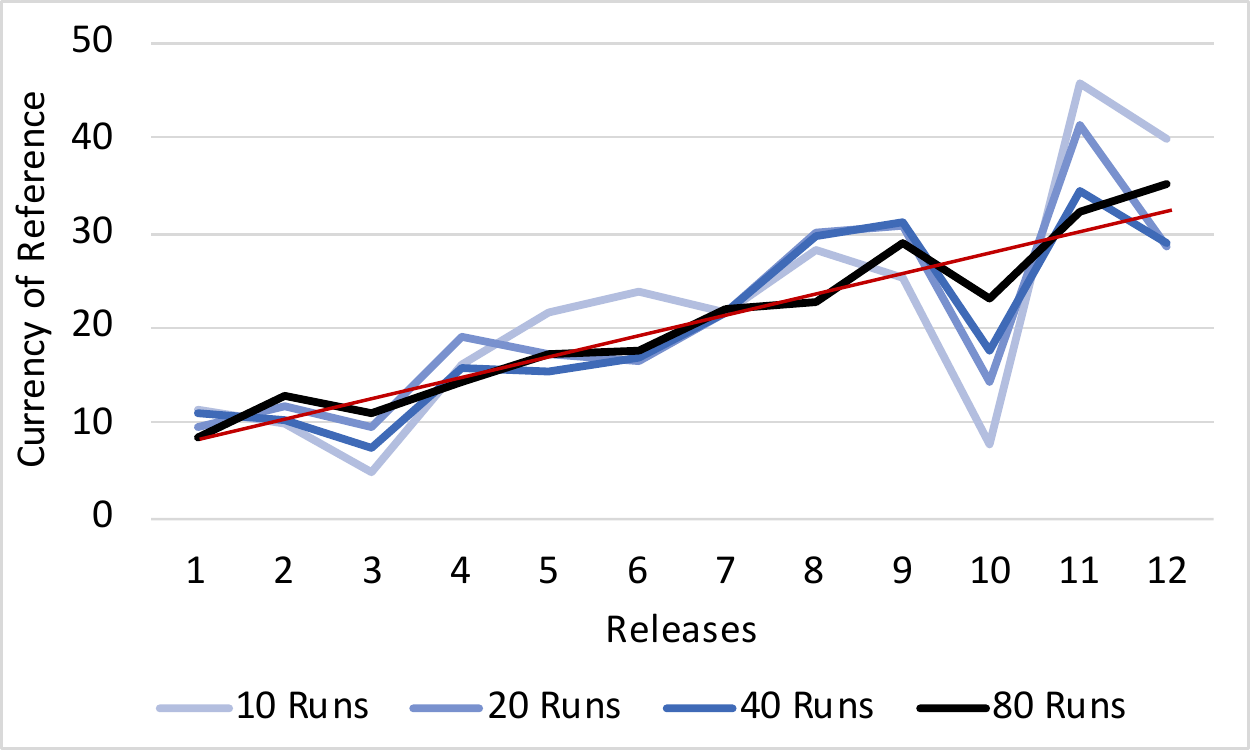}
\caption{Convergence of projected on-release migration costs of 12 releases of schema changes after 10, 20, 40, and 80 runs for the eager migration strategy.}
\label{fig:regression2}
\end{figure}

As can be noted in Figure~\ref{fig:regression}, of these 10 \MigCast{} runs relatively many values of on-release migration costs have been calculated that are much smaller than the expected correlation.\footnote{Note that, when comparing release 10 in both charts (a.) and (b.) of Figure~\ref{fig:regression}, the impact of the outlier is more pronounced on the arithmetic mean than on the median. Since external effects on the experiments have been eliminated in order to clarify the relationship between input and output, we expected a \emph{normal} distribution~\cite{Upton2002,Mackay2003,Fishman2013} and thus, decided to use the arithmetic averages in the Scatter plot charts.} Especially the sample means of releases 3, 10, and 11 would profit from more runs as their sample means deviate considerably from the expected correlation. Then, in accordance with the \emph{empirical rule}, the deviation of the sample mean to the expected value would be minimized, because the distribution of the errors in the estimates is expected to be \emph{normal}, that is, \emph{Gaussian}~\cite{Upton2002,Mackay2003,Fishman2013}. If the random values happen to be rather extreme and thus, a resulting value deviates considerably from the prospective sample mean, the subsequent resulting value is likely to be a less extreme value. This is called \emph{regression to the mean}~\cite{Upton2002,Mackay2003,Fishman2013} indicating that the sample mean eventually converges through a sufficient number of samples.% However, one should not assume that an extreme event will be evened out by another opposite extreme event, which would be an instance of the \emph{gambler's fallacy}~\cite{Tversky1971}. 

Getting back to the question as to how often the sampling has to be repeated so that the results can be generalized: We compared all metrics after multiple runs of \MigCast{} using an input parameter instantiation that has the highest variance and standard deviation, yet remains a typical scenario. This is the case with high cardinality of the relationships of the underlying data model, where single metrics vary easily by factor 100 and more. Compare the migration costs per release at a 1:25-relationship (for a detailed discussion refer to Subsection~\ref{subsec:cardinality}) for the \emph{eager} strategy after 10, 20, 40, and 80 repeated \MigCast{} runs in Figure~\ref{fig:regression2}. Because computation resources are not infinite, we need to put any gain in accuracy in perspective to the invested calculation time. The deviation from the expected correlation in this parameter configuration can be estimated to be less than 18\% after 80 runs. With most configurations the resulting values converge much quicker (<2\% deviation at 40 runs, not depicted for brevity). Therefore, we decided that 40 runs of calculation time investment into parameter configurations at low cardinality of the relationships and 80 runs for higher cardinality suffice, especially considering that in Section~\ref{sec:results}, we discuss the cumulation of the migration costs, which naturally varies much less than the migration costs per release (<2\% deviation at 80 runs).

\nop{
\begin{figure*}[ht]
\includegraphics[trim=5 5 5 0,clip,width=\textwidth]{figures/Convergence.pdf}
\caption{Convergence of cumulated migration costs after 10, 20, 40, and 80 runs of \MigCast{}; Configuration same as in Figure~\ref{fig:regression}.}
\label{fig:convergence}
\end{figure*}
}

Hence, the presented approach utilizing a Monte Carlo method of repeated sampling is justified in order to investigate how data model evolution impacts migration costs and latency while bringing the enormous complexity of data model evolution under control.

%%%%%%%%%%%%%%%%%%%%%%%%%%%%%%%%%%%%%%%%%%%%%%%%%%%%%%%
%%%%%%%%%%%%%%% SECTION 3: RELATED WORK %%%%%%%%%%%%%%% 
%%%%%%%%%%%%%%%%%%%%%%%%%%%%%%%%%%%%%%%%%%%%%%%%%%%%%%%

\section{Related Work}
\label{sec:related}

\subsubsection*{Data Model and Schema Evolution}

Research on data model and schema evolution has a long tradition, with contributions from database research, as well as software engineering research: 
There are several empirical  studies on schema evolution in real-world applications backed by relational databases~\cite{Curino2008, Skoulis2015, Vassiliadis2016,Qiu2013}. 
%This line of research has matured with the availability of public code repositories.
Various frameworks for managing schema changes in relational database management systems have been proposed, among the more recent are~\cite{Aulbach2009, Curino2013, Cleve2015, Herrmann_SIGMOD17}. Schema evolution has also been investigated in the context of XML~\cite{GuerriniMR05, BertinoGMT02} and object-oriented databases~\cite{Li1999}. 

In the context of application development with schema-flexible NoSQL database systems, new challenges arise. One factor is that such a NoSQL database system does not enforce a global schema, yet the application code will inevitably have to make assumptions about the structure of data stored in the database.
As the schema is implicitly declared within the application code,
evidence of schema changes can be observed by analyzing the change history of the application code~\cite{7884653,Scherzinger2020}, rather than an explicitly declared database schema.
Moreover, there is evidence that suggests that the application-inherent schema evolves at a higher frequency than what can be observed with schema evolution in relational database management systems~\cite{Scherzinger2020}. This makes the problem ever more pressing to deal with.

A further aggravation factor is found with zero-downtime web applications, where a new version of the application code is released against a production database that already contains data. 
On the upside, this allows great flexibility in application development, as the legacy data need not be migrated prior to a new release of the application. In fact, this is often listed as one of the sweet spots when working with NoSQL database systems. On the downside, the mismatches in the structure of legacy data and the data model expected by the application code must be actively managed.
Here, strategies for efficiently handling the migration of this legacy data takes on special importance, and is addressed next.

\subsubsection*{Data Migration}

In~\cite{EllisonCP18}, the costs, duration, and running costs are estimated for migrating entire relational database instances to the cloud, whereas we focus on the impact of legacy data on migration costs and latency caused by schema evolution. The estimates of~\cite{EllisonCP18} are based on discrete-event simulation using workload and structure models taken from logs, as well as the schema of the to-be-migrated database, whereas we investigate a range of migration scenario characteristics in a Monte Carlo approach. Since the traditional approach of \emph{eager} migration can become very expensive---especially in a cloud environment~\cite{EllisonCP18, Hillenbrand2019}---other approaches to data migration such as \emph{lazy}~\cite{Saur2016, scdm16} and proactive~\cite{Hillenbrand2019} approaches have been proposed. To our knowledge, there is no related study on the effects of the various migration strategies, comparable to ours in its systematics. In~\cite{Hillenbrand2019}, we have demoed a tool-based advisor for investigating different migration strategies. This work serves as a basis for the Monte Carlo approach presented here. 

\subsubsection*{Monte Carlo Methods in Data Management}
In terms of probabilistic methods in the context of database research, surveying uncertain data algorithms and applications, and uncertain data management has been proposed~\cite{Aggarwal2009}. In particular, Monte Carlo methods for uncertain data have been studied in detail~\cite{Haas2018, Jampani2008,Jampani2011}, also regarding the use of Monte Carlo integration methods for data clustering~\cite{Sharma2019}. 
%Probabilistic databases have been researched~\cite{Suciu2011} in depth and for instance, used for efficient query evaluation~\cite{Dalvi2007}.
% Co-Evolution - not considered here
% \cite{Lin2009}

% Note: \href causes compiling error
\nop{\begin{itemize}
    \item Cite \emph{A comparison of flexible schemas for software as a service}~\cite{Aulbach2009} - DONE
    \item References from reviews of Mark's ER2020 paper:
    \begin{itemize}
        \item Collateral Evolution of Applications and Databases~\cite{Lin2009}- NOT CONSIDERED HERE
        \item Automating the database schema evolution process~\cite{Curino2013} - DONE
        \item Understanding database schema evolution~\cite{Cleve2015} - DONE
        \item Growing up with stability: How open-source relational databases evolve~\cite{Skoulis2015} - DONE
        \item Schema Evolution and Gravitation to Rigidity: A Tale of Calmness in the Lives of Structured Data~\cite{Vassiliadis2016} - DONE
        \item \url{https://dblp.org/pers/hd/h/Herrmann_0002:Kai}{Hermann's work} wrt. evolution scenarios as sequences of SMO's - DONE
        \item A survey of schema evolution in object-oriented databases~\cite{Li1999} wrt. nesting/schema evolution of object-oriented databases - DONE
        \item \url{https://dblp.org/pers/g/Guerrini:Giovanna.html}{Guerrini's work} wrt. XML schema evolution - DONE
    \end{itemize}
    \item Data modeling in the NoSQL world~\cite{Atzeni2020} ??? - NOT CONSIDERED HERE
\end{itemize}
}

%%%%%%%%%%%%%%%%%%%%%%%%%%%%%%%%%%%%%%%%%%%%%%%%%%%%%%%
%%%%%%%%%%%%%%% SECTION 4: ARCHITECTURE %%%%%%%%%%%%%%% 
%%%%%%%%%%%%%%%%%%%%%%%%%%%%%%%%%%%%%%%%%%%%%%%%%%%%%%%

\section{Architecture and Approach}
\label{sec:architecture}

\MCMC{} runs \MigCast{} repeatedly, a tool-based advisor that we published in earlier work~\cite{Hillenbrand2019,Hillenbrand2020}. By means of repeated samplings, we contribute a systematic exploration of the search space of the scenario characteristics, thereby clarifying possible and likely impacts of schema evolution on the metrics. In this section, we discuss the architecture of the experiments, the reduction of search space complexity, and methodical aspects like the definition and verification of invariants and the reproducibility of results.

\subsection{Architecture of \MCMC{}}
\label{subsec:architecture}

\begin{figure}[t]
\centering
\includegraphics[trim=10 11 0 2,clip,width=0.475\textwidth]{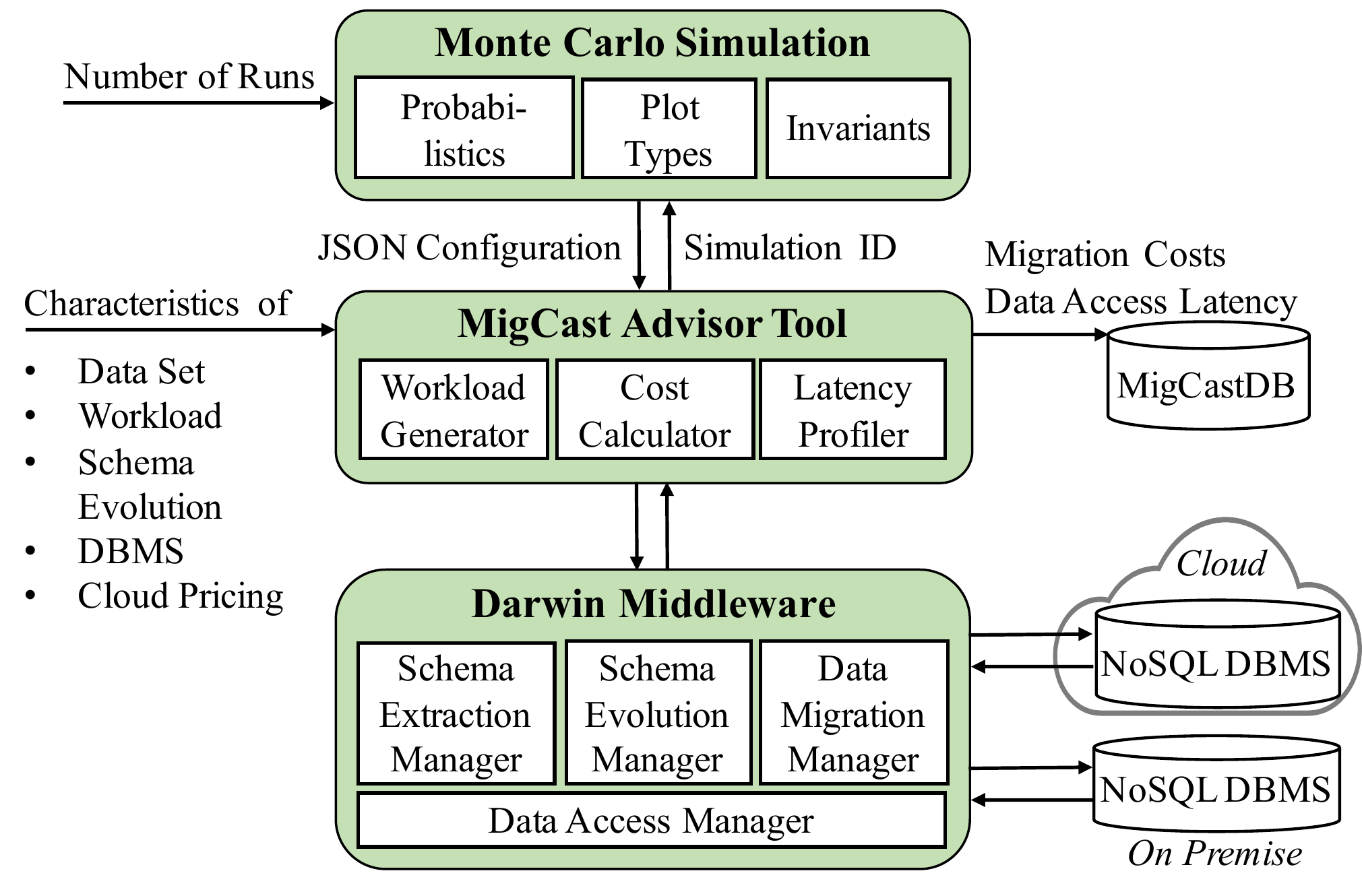}
\caption{Architecture of \MCMC{}.}
\label{fig:architecture}
\end{figure}

In Figure~\ref{fig:architecture}, an overview is depicted of the system architecture of \MCMC{}. \MigCast{} calculates the migration costs and latency in case of data model changes (modules \emph{Cost Calculator} and \emph{Latency Profiler} in Figure~\ref{fig:architecture}) based on the \Darwin{} middleware and its modules (noted here schematically, refer to~\cite{Hillenbrand2019} for further details). The calculation is based on an internal cost model that counts all I/O-requests. Latency is the time that elapses from the request until the entity is retrieved. The migration scenario characteristics that change the impact on the metrics are considered \emph{relevant} and investigated conclusively in our experiment. These characteristics are specified in the experiment as input parameters in different \MigCast{} \emph{configurations}. The investigated configurations concern the intensity and distribution of data entity accesses, the different kinds of schema changes, and the cardinality of the relationships of the underlying data model. Their instantiations and variations are summarized in Table~\ref{table:varied} of Section~\ref{sec:results}. \MCMC{} calculates repeated runs of \MigCast{} and the results are aggregated as statistical values and persisted in a dedicated database to ensure reproducibility despite the randomization.

\MigCast{} calculates the migration costs and latency for a given configuration on a case-by-case basis, taking characteristics of the data set instance, database management system, and cloud provider pricing models into account, as well as the workload of served data entity accesses and the schema changes. We refer to the data entity accesses in between two releases of schema changes as \emph{workload} (module \emph{Workload Generator} in Figure~\ref{fig:architecture}). In terms of the served \emph{workload}, we distinguish between different data access distributions and between several amounts of served workload of data entity accesses, what we refer to as \emph{workload intensity}, implemented in \MigCast{} as executions of a certain parameterizable amount of data entity accesses. The data entity accesses are randomized within the bounds of the parameterized distributions in order to facilitate the reduction in complexity by means of the Monte Carlo method. 

Regarding the characteristics of schema evolution, we differentiate between \emph{single-type} and \emph{multi-type operations} referring to the number of types that a schema modification operation (SMO) involves. By means of the migration scenario characteristic \emph{multi-type SMO complexity}, the percentage of those operations that involve more than one entity type can be determined. The concrete SMOs and the affected type(s) are then determined by randomization within the bounds of the parameterized percentages. As regards the characteristics of the data set, the \emph{cardinalities of the relationships} of the underlying data model is investigated in our experiments. These reflect typical scenarios in database applications, e.g., a 1:1-relationship in e-commerce could be a billing address of a customer, and a much higher cardinality of a relationship could be orders per customer per year. %For each \MigCast{} run and each configuration of varied input parameters, the data entity accesses (workload) and the SMOs are randomly sampled within specified bounds, yielding a complete range of possible results and statistically sound expected values. 

\begin{figure*}[t]
\centering
\includegraphics[trim=3 3 3 3,clip,width=0.9\textwidth]{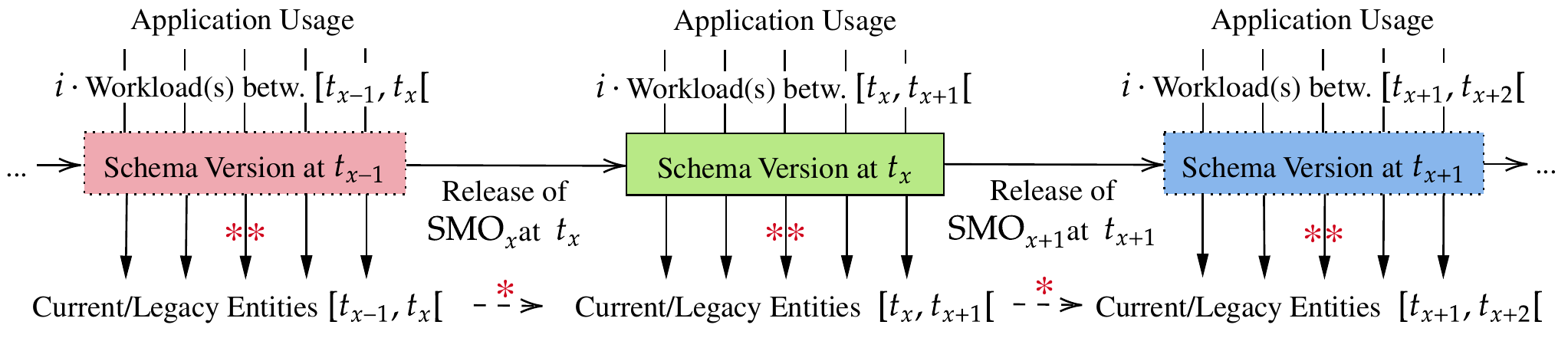}
\caption{\MigCast{} migrates legacy entities depending on the applied strategy and calculates for each release the on-read (**) and on-release (*) migration costs, which together make up the cumulated migration costs.}
\label{fig:order}
\end{figure*}

In addition to the varied parameters, the following parameters are fixed as to ensure the comparability of the investigated configurations (compare with Table~\ref{table:varied} of Section~\ref{sec:results}): The data set is determined by its \emph{real initial number of entities}, which is upscaled to its \emph{initial number of entities}. We decided for the entities to grow in number by the \emph{data growth rate} after each \emph{release of schema evolution}. An exponential growth of entities is by all means imaginable at the beginning of the project life cycle, though it would surely stabilize to a linear growth soon thereafter.  %Furthermore, the \emph{Data Set} characteristic is completed by the cardinality of the relationships of the underlying data model, which is instantiated by the data entities. 
%Depending on the particular numbers, the data model has a certain filling degree and maximal number of entities in one data model. We defined the structure to be a three-level structure, w.l.o.g.\footnote{It is also possible with our \Darwin{} middleware to upload a specific data set, including a data model and concrete set of data entities and schema evolution.} 
The amount of one workload execution is defined by the \emph{percentage of accessed data} relative to the number of existing entities in the first release, i.e., remains constant throughout the experiment. %the frequency of its executions defined in \emph{Workloads per Release} referred to as workload intensity, and its distribution parameterized in \emph{Data Access Patterns}. The characteristic \emph{Schema Evolution} is a random selection of single-type vs. multi-type schema operations in parameterizable percentages of \emph{multi-type SMO complexity} and number of \emph{Releases}. 
Lastly, the \emph{database management system (DBMS)} and \emph{price per 1M I/O-requests} are also parameterizable, yet we have based most of our discussion on relative values as opposed to absolute values.\footnote{A screenshot of the basic setting options of the \MigCast{} GUI is available at \url{https://sites.google.com/view/evolving-nosql/tools/darwin}.} %\Darwin{} interfaces with other popular NoSQL database management systems, among them \emph{MongoDB}, \emph{Couchbase}, \emph{Cassandra}, and the multi-model database \emph{ArangoDB}. 

With each release of new software, the events take place as parameterized in the following order (see Figure~\ref{fig:order}): The specified amounts of randomized workload of entity accesses are served, which can cause on-read migration costs with migration strategies that delay migrating legacy entities (** in Figure~\ref{fig:order}). Then, an SMO is applied, which is randomly chosen according to the specified multi-type SMO complexity. In case of the \emph{eager}, \emph{incremental}, and \emph{predictive} strategies, certain amounts of affected legacy entities are migrated then causing on-release migration costs (* in Figure~\ref{fig:order}). %Then the \emph{Migration Debt} is calculated. 
Lastly in each release, new data entities are generated, supplementing the existing data set according to the specified cardinalities of the relationships of the underlying data model.

\subsection{Reducing the Search Space Complexity} 
\label{subsec:searchspace}

Although the computational effort is reduced by the randomization of data entity accesses through the Monte Carlo method, one \MigCast{} run for an average configuration and a sampling size of 1,000 entities took approx. 4.5 minutes on our server. Using a typical number of data entities would far exceed the computational resources. Thus, we limited the sampling size of the experiment, yet still ascertaining that the results are representative. We ran the experiments with different sampling sizes and eventually decided for 1,000 entities, since the resulting metrics of the migration strategies and their variance do not significantly change with more entities. We then multiply the results with a factor of 10,000 to scale up the resulting metrics to represent a typical amount of 10M data entities. 

Furthermore, the decision on the number of parameter variations of the investigated configurations has to be put into perspective in terms of the computational effort of one such cost calculation. The problem search space has exponential complexity and  thus, calculating the metrics for all combinations of migration scenario characteristics would by far exceed our life times. We decided to focus on 90 different typical configurations that cover the search space and allow uncovering the correlation between the scenario characteristics and their impacts on the metrics. We discussed in Section~\ref{sec:preliminaries} how many runs of the cost calculations for each configuration are necessary for the metrics to converge. 

% All these decisions regarding the experiment are the result of a process with the goal of making substantial observations despite the huge problem space. %The results are discussed in the following section, where the concrete instantiations of the input parameters are summarized in Table~\ref{table:varied}.

\subsection{Methodical Aspects}% of \MCMC{}}
\label{subsec:evaluation}
\label{subsec:repeatability}

In order to methodically approach verifying our hypotheses, we formulated 29 invariants how we expected the migration strategies to perform in terms of migration costs and latency in each of the investigated scenarios. As a welcome side effect, we could check and ascertain model consistency and validate the implementation. All charts of all \MigCast{} runs are routinely evaluated by the invariants.\footnote{See a screenshot of analyzed invariants available at \url{https://sites.google.com/view/evolving-nosql/tools/darwin}.} A traffic light system facilitates a relatively quick assessment of whether the results fulfill the invariants, especially quick compared to manually checking 90 configurations times 40/80 runs of logs.

The invariants can be distinguished as either being \emph{requirements} or \emph{hypotheses}. The former have to be met in order to prove consistency, shown by a green check mark, or if not met, by a red X mark. In order to pursue our hypotheses how the input parameters correlate with the results, invariants have been formulated as \emph{tendencies} that, if met, support our hypotheses, then indicated by green check mark as well. In case that a \emph{tendency} is not met at every single release, a yellow X mark indicates this. The complete, persisted logs can be accessed by means of the configuration number and run index in order to investigate whether the invariant is just temporarily not met, e.g., through outliers, or whether it contradicts a hypothesis. The results presented in Section~\ref{sec:results} are all confirmed through the invariants.

The second method that we used to ascertain model consistency addresses the circumstance that a Monte Carlo approach is probabilistic in nature. Despite including randomly instantiated variables, the experiments have to be reproducible in order to meet certain scientific requirements known as \emph{data provenance}~\cite{Herschel2017}. A MongoDB database, \emph{MigCastDB} in Figure~\ref{fig:architecture}, persists all results and calculated statistical measures in order to ensure their reproducibility, understandability, and comparability.\footnote{Be referred to our project website for a complete data model of \emph{MigCastDB} and a scrollshot of all invariants: \url{https://sites.google.com/view/evolving-nosql/tools/darwin}.} %The data model was created using a \emph{Schema Visualizer} tool designed by our co-author Shamil Nabiyev, see \url{https://demo.schemavisualizer.com/}.} %Thereby, we can reproduce any experiment using the same randomly sampled SMOs and data accesses by specifying the experiment's \emph{ObjectID}. 
Note that, when repeating an experiment, the latency is the only metric that can vary, although in our setting we minimized external effects on the experiment by removing all potentially interfering routines on the server. In addition, the repeated sampling keeps any effects on latency low. %We can deduce that, in essence, the presented experiment necessarily uncovers the correlation of both metrics and a Monte Carlo method is thus justified. 
Last but not least, we ran the \MigCast{} tool once after starting the DBMS in order to ensure the comparability of all runs.

%%%%%%%%%%%%%%%%%%%%%%%%%%%%%%%%%%%%%%%%%%%%%%%%%%%%%%%
%%%%%%%%%%%%%%% SECTION 5: RESULTS %%%%%%%%%%%%%%%%%%%% 
%%%%%%%%%%%%%%%%%%%%%%%%%%%%%%%%%%%%%%%%%%%%%%%%%%%%%%%

\section{Putting the Results into Context}
\label{sec:results}

Legacy entities originate in schema changes of the underlying data model. Migration strategies handle legacy entities differently, thus a particular strategy may be more suitable in one migration scenario but less so in another. By means of \MCMC{}, we are able to clarify how migration scenario characteristics, represented by instantiations of \MigCast{}'s input parameters, impact the competing metrics migration costs and latency. The knowledge of this correlation put stakeholders of a software project in the position to remain in control of the operative costs for data model evolution and base their decisions during software application development on transparency regarding the impact on the metrics. %, for instance, by changing the migration strategy or the release strategy to suit new production settings.

In Subsection~\ref{subsec:architecture} we have outlined the architecture of \MCMC{}. The concrete instantiations of the input parameters can be divided into fixed and varied parameters, the latter of which we investigated (see Table~\ref{table:varied}): Two workload distribution patterns (Pareto, uniform), three degrees of workload intensity (low, medium, and high), five multi-type SMO complexity variations (steps of 25 percentage points), and three cardinalities of 1:n-relationships (n being 1, 10, or 25). The served workload of data entity accesses as well as the concrete SMOs are randomized within the bounds of these parameter instantiations. These combinations result in 90 different \MigCast{} configurations amounting to 3,600 runs and approx. 270h of computation time on a server running Ubuntu 16.04.6 and MongoDB (Server: 4.0.2, Driver: 3.9.0) on 24 CPUs and 128 GB RAM.\footnote{We recorded the application data metrics through the Grafana tool. The charts of CPU load, processes, and used memory can be viewed at \url{https://sites.google.com/view/evolving-nosql/tools/darwin}.} %Table~\ref{table:varied} summarizes the input parameter instantiations and variations of \MCMC{}.

\begin{table}[t]
\begin{center}\begin{footnotesize}
\begin{tabularx}{0.475\textwidth}{lll} \toprule
\emph{Parameter Category/Name} & \emph{Default} & \emph{Investigated Variations} \\\toprule
\emph{Data Set} & & \\\midrule
Initial Number of Entities& \multicolumn{2}{l}{10M (upscaled)}  \\
Real Initial Number of Entities & 1,000 &  \\
Data Growth Rate (per Release) & 10\% & \\ 
Cardinality of 1:n-Relationships & 1:1 & 1:1, 1:10, 1:25 \\\midrule
\emph{Workload} & & \\\midrule
Distribution of Data Accesses & Pareto 80/20 & Pareto 80/20, Uniform \\
Intensity, in measures of: & Med: 2x 10\% & Low (1x), Med (2x), High (4x)\\
Percentage Accessed Data & 10\% (const.) & \\\midrule
\emph{Schema Evolution} & &  \\\midrule
Releases & 12 & \\
Multi-type SMO complexity & 25\% & 0\%, 25\%, 50\%, 75\%, 100\% \\\midrule
\emph{DBMS and Cloud Pricing} & & \\\midrule
DBMS & MongoDB & \\
Price per 1M I/O Requests & USD 0.2 & \\
\bottomrule
\end{tabularx}
\end{footnotesize}
\end{center}
\caption{Parameter instantiations of \MCMC{}.}
\label{table:varied}
\end{table}

\begin{figure}[h]
\centering
\includegraphics[trim=0 0 0 8,clip,width=0.35\textwidth]{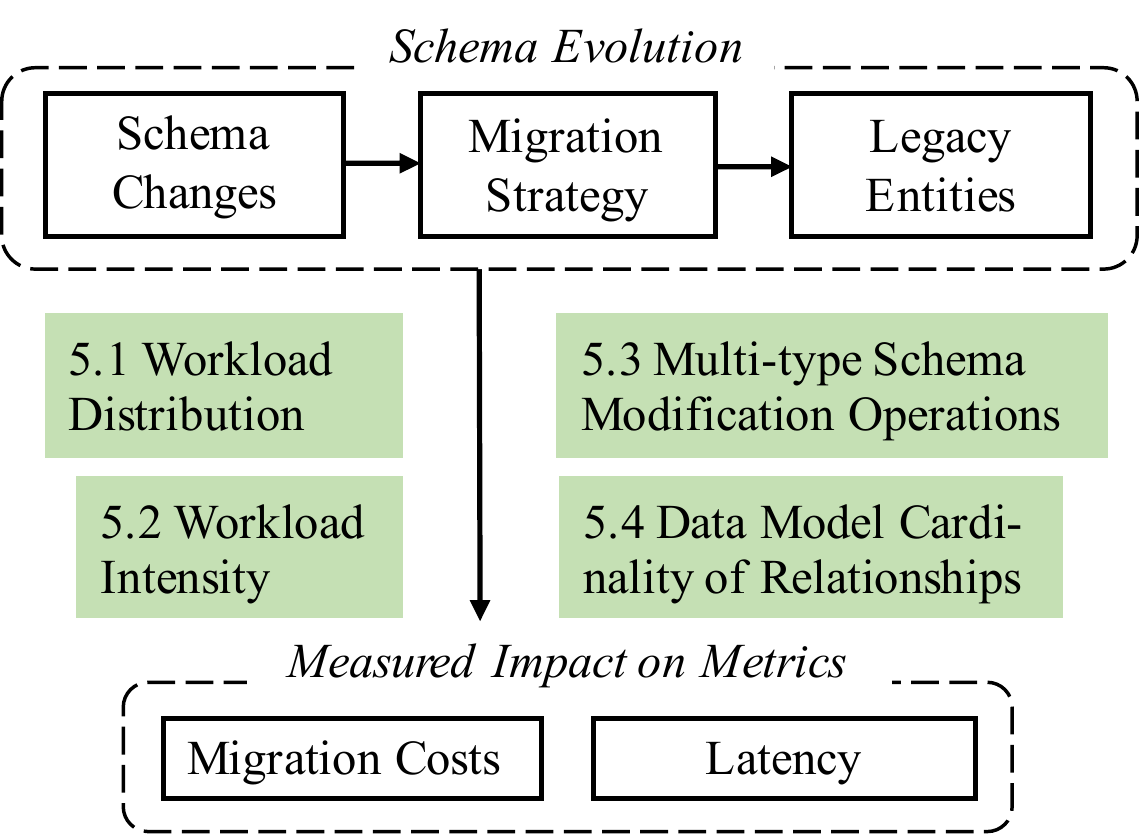}
\caption{Measuring the impact of schema evolution with respect to relevant migration scenario characteristics.}
\label{fig:overview}
\end{figure}

In the following, we present the results of the \MCMC{} experiments. We categorize them into four subsections corresponding to the investigated scenario characteristics (Figure~\ref{fig:overview}). Some of them amplify the differences that exist between the different migration strategies in respect of a particular metric, and others level these differences while varying the investigated characteristic. The graphs depicting a metric of the strategies then diverge more or less depending on the variation of an input parameter.

We start in Subsection~\ref{subsec:concentration} with the analysis how the distribution of the served workload of data entity accesses influences the measured impact on the metrics for each of different migration strategies. In Subsection~\ref{subsec:workload} we discuss the impact in respect of varying intensity of workload between releases of schema changes. In Subsection~\ref{subsec:complexity}, the impact is analyzed in respect of how many entity types the data model changes affect, and in Subsection~\ref{subsec:cardinality} the influence of the cardinality of the relationships of the underlying data model is investigated. Note that the diagrams of the cumulated migration costs of each scenario characteristic variation (Figures~\ref{fig:distribution2},~\ref{fig:workload2}~\ref{fig:complexity0-75MCC}, and~\ref{fig:card}), which accompany each subsection, refer to \emph{relative} costs so that each scenario characteristic can be easily compared with the others (the most influencing charactertistic shows 100\% on the y-axis, the others 60\% each).

\subsection{Concentration of Data Entity Accesses}
\label{subsec:concentration}

\begin{figure*}[ht]
\centering
\includegraphics[trim=5 7 26 3,clip,width=\textwidth]{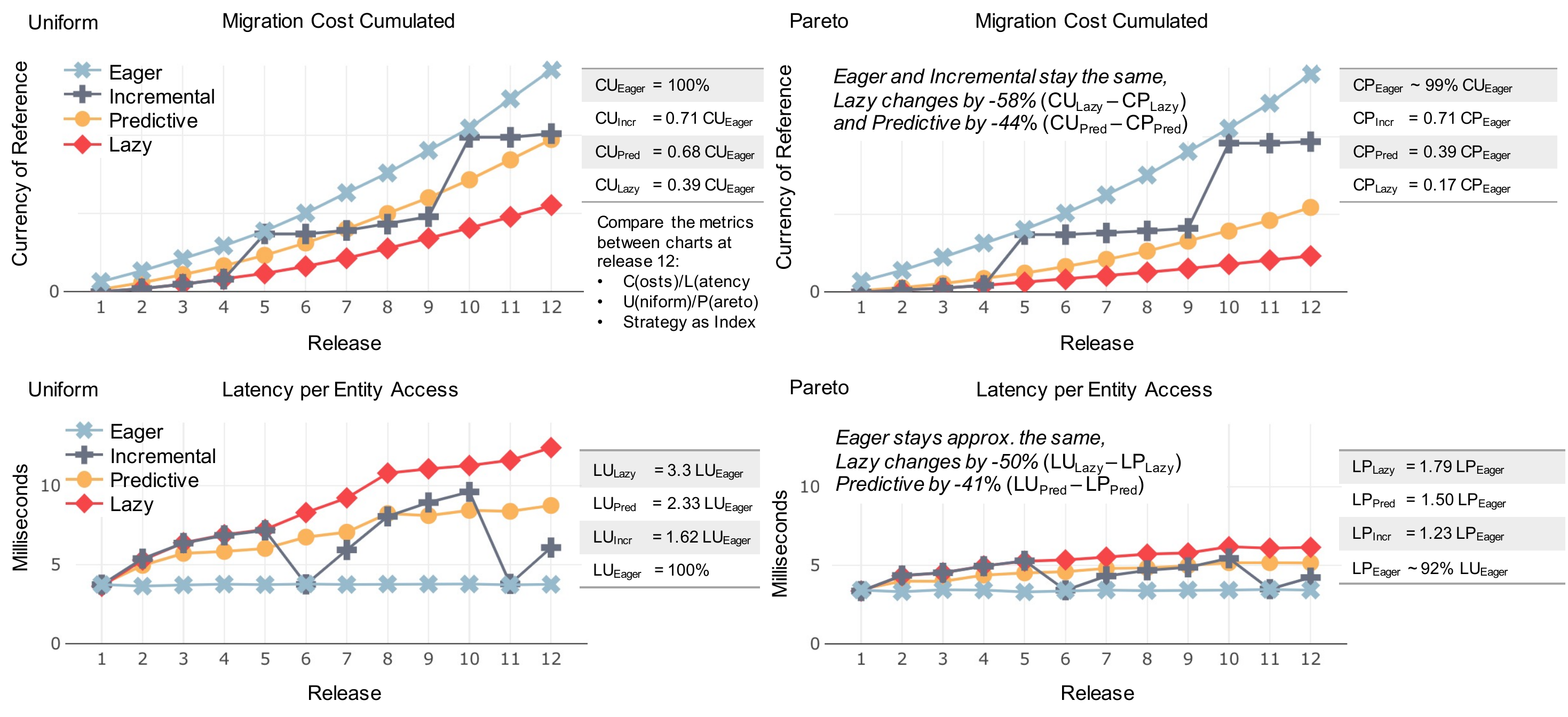}
\caption{Comparing the uniform with the Pareto distribution: The migration strategy graphs diverge more w.r.t. cumulated migration costs and less w.r.t. latency; Configuration: Medium workload, 25\% multi-type SMO complexity, standard cardinality.}
\label{fig:distribution}
\end{figure*}

\begin{figure}[ht]
\centering
\includegraphics[trim=12 40 13 6,clip,width=0.41\textwidth]{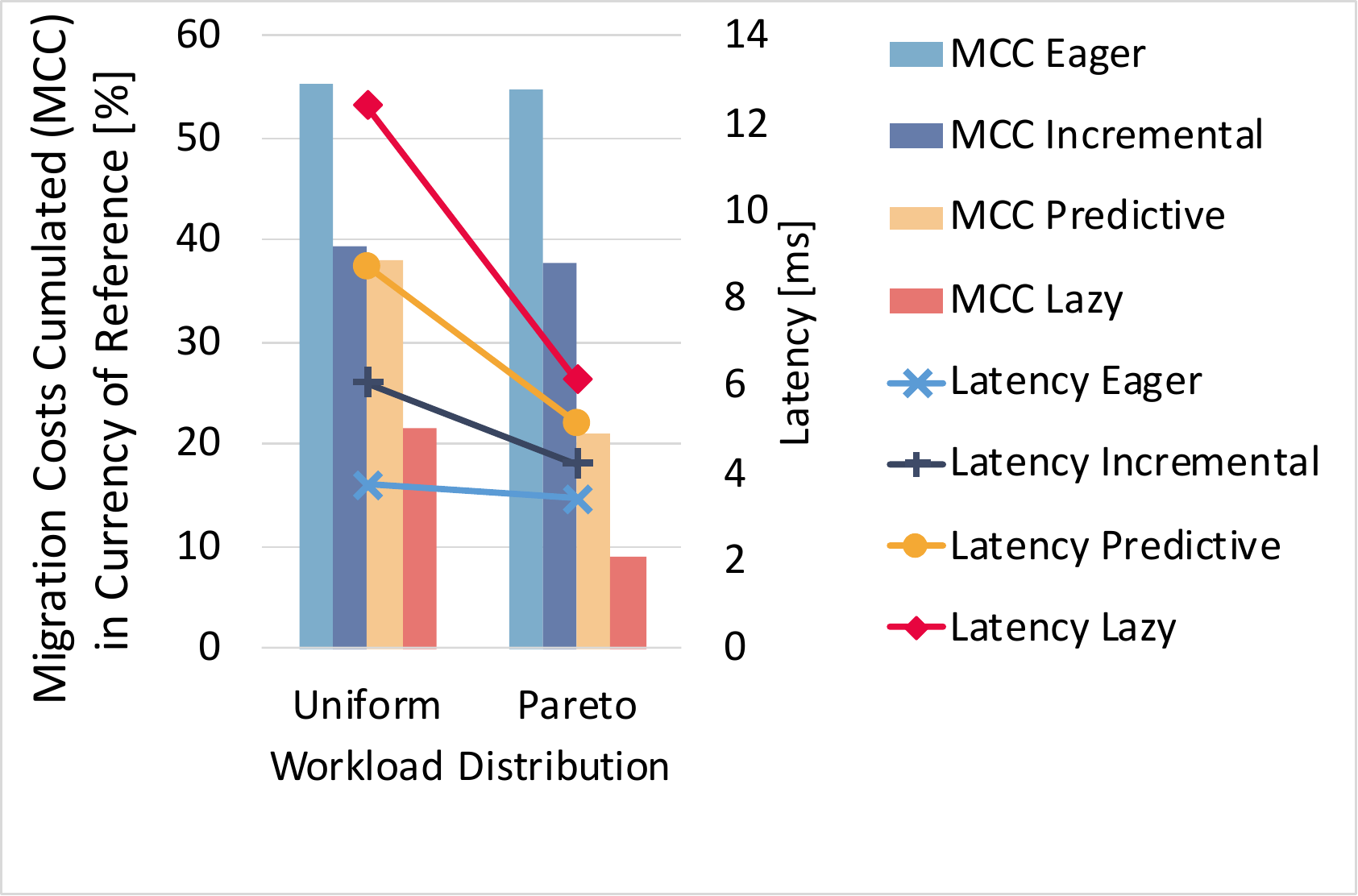}
\caption{Comparing uniform and Pareto distributions after 12 releases; Configuration same as in Figure~\ref{fig:distribution}.}
\label{fig:distribution2}
\end{figure}

We begin with the investigation on how the distribution of entity accesses of the workload impacts the resulting metrics of migration costs and latency. We ran \MCMC{} with the \emph{uniform} and with the \emph{Pareto} distribution, the resulting metrics of which are depicted in Figures~\ref{fig:distribution} and~\ref{fig:distribution2}. The first figure shows the graphs indicating how the metrics develop with the releases of schema changes, of which the second figure is a compact summary after 12 releases. With the \emph{uniform} distribution the probability of one single entity being accessed is the same for all entities, whereas with the Pareto distribution the probability is greater with \emph{hot} data entities which are accessed more frequently. We used a Pareto pattern that is common in OLTP database applications~\cite{DBLP:conf/icde/LevandoskiLS13}, where 80\% of the workload accesses concentrate on 20\% of the data entities, the \emph{hot} data, and 20\% of the workload accesses are distributed among 80\% of the data entities, the \emph{cold} data. 

As can be seen in Figure~\ref{fig:distribution}, the \emph{lazy} migration strategy performs better under the Pareto distribution both in terms of the metrics migration costs and latency compared to the uniform distribution. With each distribution pattern any one entity is possibly accessed repeatedly, which is much more probable with the Pareto distribution where the workload concentrates on hot data. Specifically, under the Pareto distribution the migration costs drop by 58\% and latency drops by 50\% for the \emph{lazy} migration strategy.

%For each configuration of the \MCMC{} simulation the selection of accessed entities was random within the constraint specified in the parameter \emph{Data Access Pattern}. 
%With each distribution pattern any one entity is possibly accessed repeatedly, which is more or less probable. In Figure~\ref{fig:pareto-uniform}, the number of accesses per entity is visualized for the uniform and for the Pareto distribution, calculated for the \emph{Percentage Accessed Data} of 10\%. As the dependent variable in each diagram, the distribution of the percentage of accessed entities is plotted against the amounts of served entity accesses in between two releases of schema evolution as discrete values of workload executions. When comparing the two distribution patterns it becomes apparent how under the Pareto distribution the accesses concentrate on hot data while the percentage of entities that are not being accessed at all (green bars) is higher than under the uniform distribution. The entities that are being accessed under the Pareto distribution are about half of the accessed entities under the uniform distribution. This can be recognized in~\ref{fig:distribution}, where the migration costs change by -59\% under the Pareto distribution, and latency still changes by -36\% for the \emph{lazy} migration strategy.  %Refer to the following subsection~\ref{subsec:workload} for a detailed discussion how workload affects the metrics. 

\nop{
\begin{figure}[ht]
\centering
\includegraphics[trim=2 2 2 2,clip,width=0.475\textwidth]{figures/Pareto-uniform.pdf}
\caption{Comparing uniform and Pareto distribution w.r.t. the concentration of data entity accesses.}
\label{fig:pareto-uniform}
\end{figure}
}

The distribution of entity accesses has an influence on the metrics due to the varying number and age of the legacy entities. The heterogeneity of the data in terms of schema versioning not only depends on the distribution of entity accesses, but also on the chosen migration strategy. For this discussion it has been proven useful to differentiate between on-release and on-read migration costs, which together make up the cumulated migration costs. On-release migration costs depend on how many entities are affected by a schema change and how these legacy entities are being handled. With the \emph{eager} migration strategy, all legacy entities are migrated to the current schema version, so that no runtime overhead exists. In this case, only on-release migration costs contribute to the cumulated migration costs. % as no further on-read migration costs are caused---until the next release. 
While latency is minimal for \emph{eager} migration, this approach also causes maximal migration costs compared to other migration strategies. %The eager approach represents one stance that can be taken with regard to the tradeoff between migration costs and latency that has to be decided upon.
In contrast to on-release migration costs, on-read migration costs are caused when entities are being accessed that exist in older versions than the current schema indicates. With the \emph{lazy} migration strategy, legacy data remains unchanged in the event of releases of data model changes. %When legacy entities are being queried, the query is rewritten according to the schema information and pertinent legacy entities are migrated on-the-fly in order to adhere to the new data model, which causes a substantial runtime overhead~\cite{Hillenbrand2019,Saur2016}.
Savings in terms of on-release migration costs are paid for in form of on-read migration costs and a higher latency of the \emph{lazy} strategy, once a legacy entity needs to be accessed. As can be seen in Figures~\ref{fig:distribution} and~\ref{fig:distribution2} for both distributions, the \emph{lazy} strategy represents an upper bound and the \emph{eager} strategy a lower bound with respect to the cumulated migration costs, and vice versa with respect to latency. %Compare this with Figure~\ref{fig:distribution2}, in which both metrics are plotted against the distributions, the line chart depicting the latency, and the bar chart depicting the cumulated migration costs after 12 releases of schema changes.

Regarding the distribution of accesses it can be observed that the \emph{lazy} migration strategy particularly benefits in terms of both metrics when entity accesses concentrate on hot data. Once a legacy entity is migrated and then up-to-date, future queries benefit from the update in terms of lower on-read migration costs and latency, which pays off in case of repeated accesses on the same entities. In our experiments, 58\% of the cumulated migration costs can be saved under the Pareto distribution. Compared to the \emph{eager} strategy, the \emph{lazy} strategy causes 39\% of the costs under the uniform distribution and only 17\% under the Pareto distribution (see Figure~\ref{fig:distribution}).

Located in between the \emph{lazy} and \emph{eager} approaches is the \emph{incremental} migration, which migrates all legacy entities at certain preset points in time, here at releases 5 and 10. Characteristic for \emph{incremental} strategy is that it fluctuates between the \emph{eager} and \emph{lazy} strategy both in terms of migration costs and latency. This can be utilized when the workload to be served is known to vary in a certain recurring pattern, to which the \emph{increments} of migrating legacy entities can be adapted. Lazy periods of time are then interrupted by regular bouts of tidying up the structurally heterogeneous database instance in order to get rid of the runtime overhead caused by updating legacy entities on-the-fly when being accessed.\footnote{Note that, due to this oscillation, the snapshots of the metrics after 12 releases as in Figure~\ref{fig:distribution2}, may not always fully represent the graphs in Figure~\ref{fig:distribution} as an average value would do. We have considered this in our evaluation.}

The \emph{predictive} migration strategy shows its full potential under the Pareto distribution, benefiting from the presence of hot data in similar measure like the \emph{lazy} strategy: While latency decreases by 41\%, 44\% of migration costs can be saved (comparing the uniform with the Pareto distribution), latency decreases by 50\% and migration costs by 58\% with the \emph{lazy} strategy (refer to Figures~\ref{fig:distribution} and~\ref{fig:distribution2}). Comparing the \emph{predictive} with the \emph{incremental} strategy, it should be noted that the predictive strategy convinces with more stable and thus more predictable metrics, especially important for SLA-compliance when the migration increments cannot be matched to intervals of low workload. Furthermore, the \emph{predictive} strategy can make use of the Pareto distribution and saves more migration costs. %Not depicted but for the sake of completeness, under the \emph{Pareto} distribution migration debt remains approx. the same with the \emph{lazy} strategy but increases by 28\%, as there are nominally more cold data entities that were not needed to be migrated.

%Furthermore, we have implemented an \emph{adaptive} migration strategy (turquoise graphs in all charts) in our schema management middleware \emph{Darwin}~\cite{Hillenbrand2019}, which improves on the \emph{predictive} approach by increasing the prediction set size when many schema modification operations have occurred. This prevents a latency peak when legacy entities are being accessed that have been affected by multiple data model changes. For example, extensive schema evolution usually happens during major schema changes in agile development settings. The situation of a backlog of data model changes is then prevented by adapting the prediction set size in case that a certain number of multi-type operations have accrued. We have described \emph{adaptive} migration more in detail in Section~\ref{sec:architecture}. With \emph{predictive} and \emph{adaptive} migration, latency gets more stable, which is especially important for application performance as peaks of extremely long latency times are systematically avoided. In terms of the distribution of entity accesses, the \emph{adaptive} migration strategy responds just like the \emph{predictive} approach as it benefits from the presence of hot data. Since our simulation levels out the results through its Monte Carlo approach, the difference between \emph{predictive} and \emph{adaptive} approaches are merely subtle (the yellow graphs hide behind the turquoise graphs), however, in subsection~\ref{subsec:complexity} a scenario is discussed where \emph{adaptive} migration is able to show its advantages.

Interpreting these findings differently, if the hot data of an established Pareto pattern changes to a different set of hot data, then higher migration costs and latency must be expected intermittently, up to the migration costs and latency found for the uniform distribution. The metrics then converge again to previously measured values once this new Pareto pattern is established.

Summing up, if the workload distribution exhibits a pattern in which entity accesses concentrate on hot data compared to randomly distributed entity accesses, then migration strategies that delay migrating legacy entities in the event of schema evolution perform better in terms of a lower latency. At the same time, migration strategies that utilize the Pareto principle can save more migration costs, while the \emph{eager} strategy remains the same and the \emph{incremental} strategy saves costs very slightly on average. While latency increases with the releases of schema changes for strategies that utilize the Pareto principle, it is generally more stable with regard to all scenario characteristics than with the \emph{incremental} strategy. In general, the concentration of data entity accesses amplifies the differences of the migration strategies in terms of migration costs and levels the differences in terms of latency. The opposite is true if the data entity accesses do not concentrate on hot data, i.e., if the probability of being accessed is the same for all data entities, then the differences are amplified in terms of latency and leveled in terms of migration costs. %, and remains neutral in terms of migration debt with the exception of the predictive strategy that allows higher debt.

\begin{observation}
In terms of the distribution of entity accesses, the \emph{eager} strategy remains invariant and represents an upper bound with respect to cumulated migration costs and a lower bound with respect to latency, and vice versa for the \emph{lazy} strategy in terms of upper/lower bounds. % and migration debt.
It also holds that under the Pareto distribution compared to the uniform distribution:
\begin{compactitem}
\item The cumulated migration costs diverge more, thus more costs can be saved with strategies utilizing the Pareto principle, i.e., the \emph{lazy} and \emph{predictive} strategies, and (very slightly) with the \emph{incremental} strategy.
\item Across all strategies, latency diverges less when accesses concentrate on hot entities.
%\item Migration debt remains approx. the same when accesses concentrate on hot entities, except for the \emph{predictive} strategy where debt increases considerably.
\item The opposite of the two previous points holds true under the uniform distribution compared to the Pareto distribution.
\end{compactitem}
\end{observation}

\subsection{Workload Intensity}
\label{subsec:workload}

\begin{figure}[ht] 
\centering
\includegraphics[trim=12 7 8 8,clip,width=0.475\textwidth]{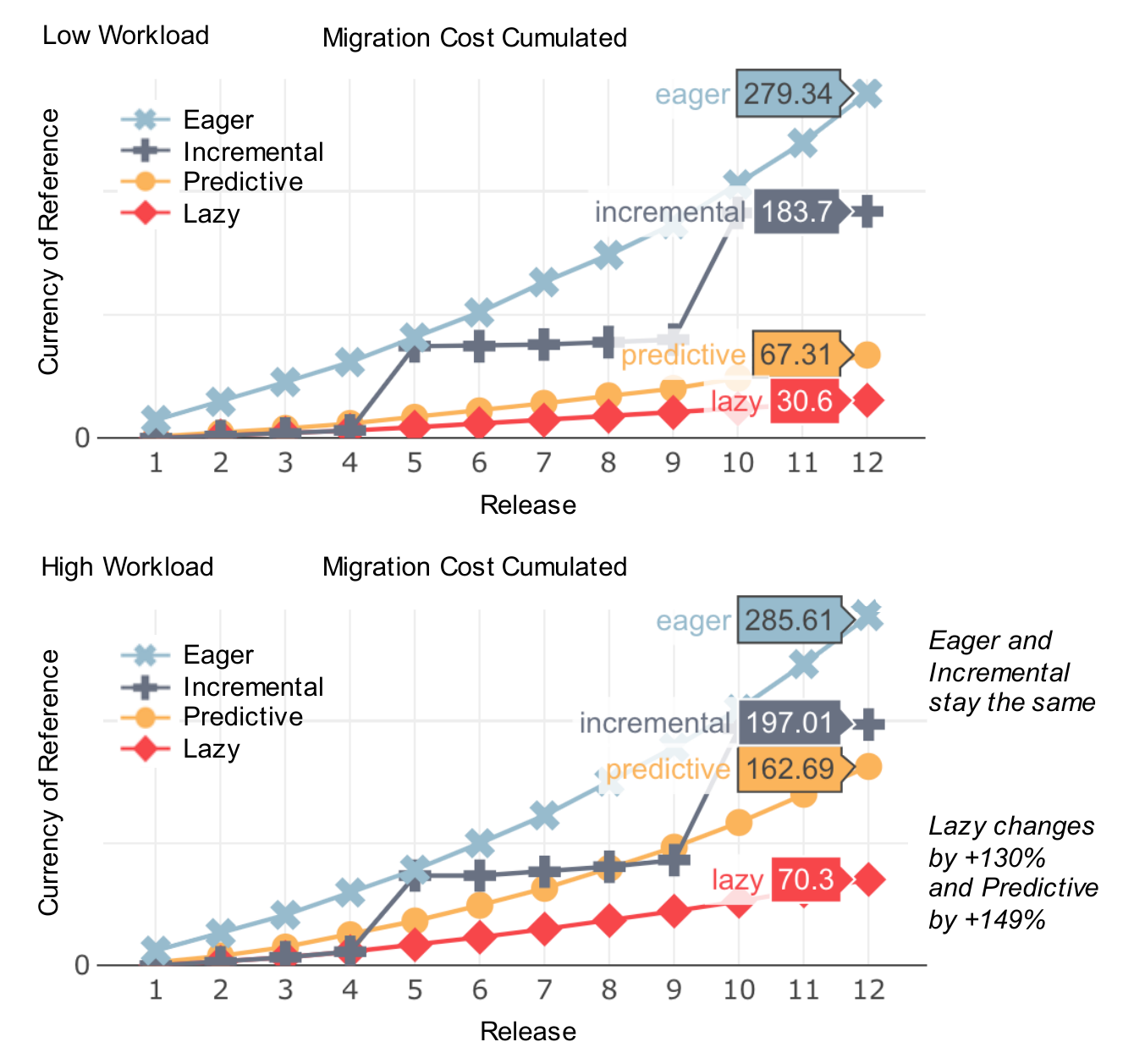}
\caption{With increasing workload, the graphs indicating the cumulated migration costs for different strategies diverge less; Configuration: Pareto-distributed workload, 25\% multi-type SMO complexity, standard cardinality.}
\label{fig:workload}
\end{figure}

In order to investigate how workload influences the metrics, we have run \MCMC{} with varying intensity of workload between releases of schema changes. The charts in Figure~\ref{fig:workload} show the migration costs for workload intensity from low to high, where low equals one workload execution and high equals four times as much, simulating varying intensity of served entity accesses in between releases of schema changes. Each workload execution is a certain amount of entity accesses, in our experiments 10\%, relative to the initial number of entities in the first release. The amounts of this percentage of accessed data remains the same throughout the releases despite the growth of data, simulating a constant workload over time. Each entity access is random within the bounds of the Pareto distribution, which is the default workload distribution. %With increasing workload the entities that are being accessed concentrate even more, especially under the Pareto distribution,

%Generally, in \MCMC{} the number of accesses depends on the following simulation parameters: \emph{Real Initial Number of Entities} specified as 1000, \emph{Percentage Accessed Data} is 10\%, and \emph{Data Growth Rate} also 10\%, if not stated otherwise. The results are then projected to a realistic \emph{Initial Number of Entities} of $10M$ entities. As to the reasons why this parameterization is justified, refer to Section~\ref{sec:architecture}. As explained, we can see in Figure~\ref{fig:workload} that the cumulated migration costs of the \emph{eager} strategy remain the same with increasing workload. After 40 runs of the \MCMC{} simulation the costs for \emph{eager} have leveled out at their limits when comparing one, two and four workloads.

\begin{figure}[ht] 
\centering
\includegraphics[trim=11 51 6 6,clip,width=0.475\textwidth]{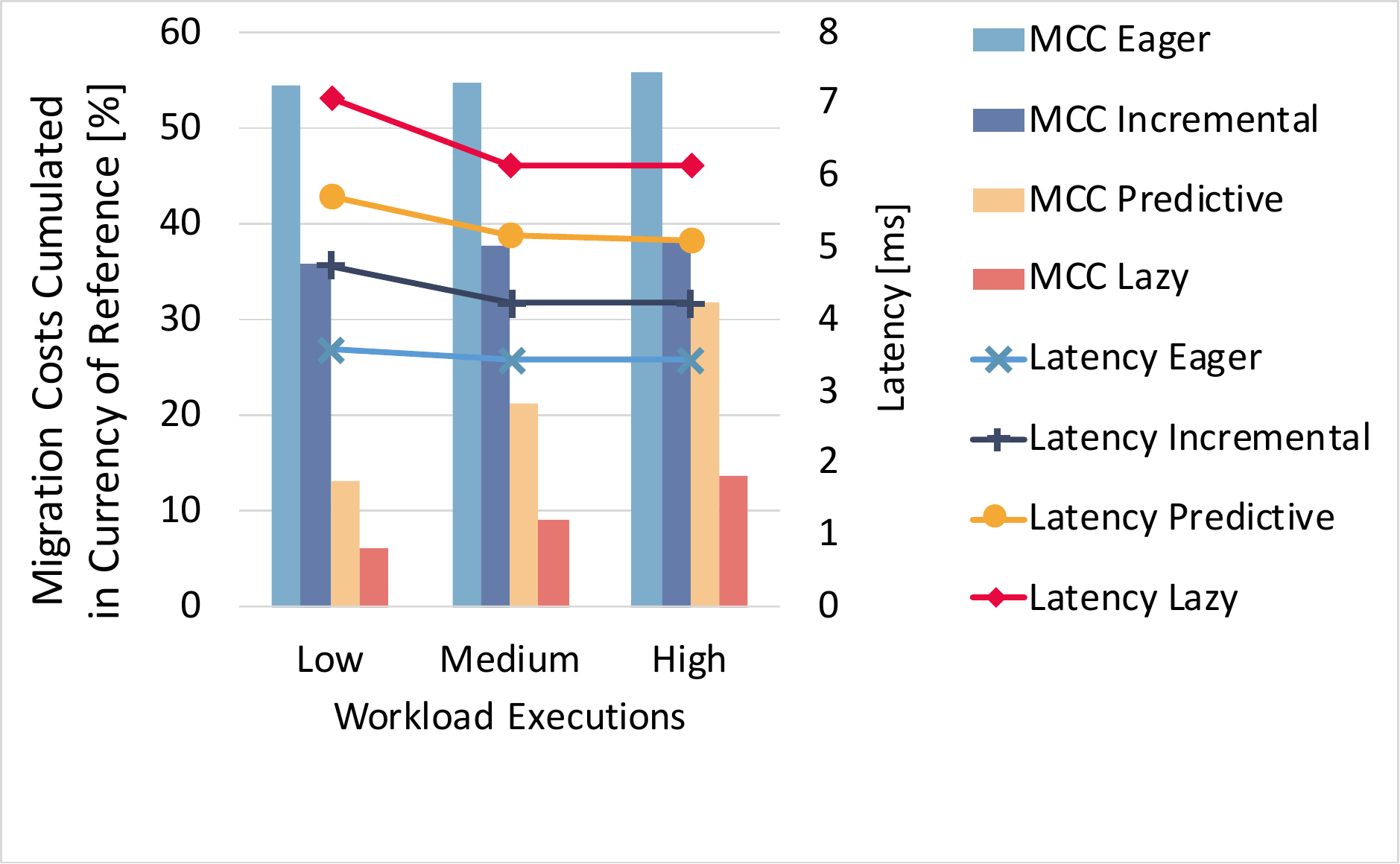}
\caption{With increasing workload, cumulated migration costs increase and latency decreases on average; Configuration same as in Figure~\ref{fig:workload}.}
\label{fig:workload2}
\end{figure}

As explained earlier, we can see in Figures~\ref{fig:workload} and~\ref{fig:workload2} that the metrics of the \emph{eager} strategy remain approximately the same with increasing workload. In contrast to on-release migration costs that are caused by releasing data model changes, on-read migration costs are caused when entities are being accessed that abide by older versions than the current schema indicates. Now, with increasing workload, the percentage of legacy entities declines for the \emph{lazy}, \emph{incremental}, and \emph{predictive} strategies and thus, it can be observed in Figure~\ref{fig:workload} that, while the cumulated migration costs continuously increase with each release, the advantage over other strategies diminishes, and the graphs diverge less. The proactive strategies, which act in advance of situations when migrating legacy entities could cause latency overhead, are consistently located between the \emph{lazy} and the \emph{eager} strategies for both metrics depending on their investment into the structural homogeneity of the database. %Regarding the latency, it can be observed that although the increase of the amount to medium workload notable in Figure~\ref{fig:workload2} for strategies that delay migrating legacy entities, does not yet convert into a decrease of lower latency, this expected effect is notable with a high amount of workload, which twice the medium amount of workload. 

It can be observed in Figure~\ref{fig:workload2} regarding the latency that an increase of the workloads corresponds to a decrease of latency for strategies that delay migrating legacy entities. Specifically, it can be noted that with four times the amount of workload (low to high) the migration costs for the \emph{lazy} strategy increase by 130\% (+14 percentage points in respect of the charges for \emph{eager}) and for the \emph{predictive} strategy by 149\% (+44 percentage points), refer to Figure~\ref{fig:workload}. Latency then drops for \emph{lazy} from 7.09ms to 6.15ms, that is by 13\% (-8 percentage points in respect of the latency for \emph{eager}), and for \emph{predictive} from 5.72ms to 5.11ms, that is by 11\% (-7 percentage points), refer to Figure~\ref{fig:workload2}. 

This shows a particularly efficient investment of migration costs into hot data in case of low workload and for strategies that utilize the Pareto principle. This wears off with increasing workload, then bringing those strategies closer to the \emph{incremental} strategy. The reason that the \emph{predictive} strategy gets closer to the \emph{incremental} strategy with higher workload can be found in the choice of the prediction set size of the \emph{predictive} strategy: If the workload is lower (1/2/4 times 10\% in our experiment) than the prediction set size (30\%), then the metrics for the \emph{predictive} strategy are closer to those for the \emph{lazy} strategy. In case that the whole extension of a relatively high prediction set size is used, the metrics get closer to the \emph{incremental} strategy. The efficiency of migration cost investment for the \emph{predictive} strategy thus depends on the workload and on the prediction set size. Interestingly, with the chosen size of the prediction set and thereby the investment of migration costs, the resulting latency of the \emph{predictive} strategy proportionally fits in relative to the other strategies. A detailed investigation on the correlation of different prediction set sizes and resulting metrics is part of our future research, following up on~\cite{Hillenbrand2020}.

In summary, we can state that the graphs of the different migration strategies are less divergent with higher workload in respect of migration costs, because more entity accesses result in higher migration costs, except for \emph{eager} migration which stays invariant under the workload. Latency also becomes less divergent as more legacy entities are migrated and up-to-date with strategies that delay migrating legacy entities. Again, \emph{eager} migration stays invariant under workload for latency as well. The differences between the strategies can be observed as more or less proportional in terms of their investment into the structural homogeneity of the database and the resulting improvement in latency. %However, counter-intuitively, migration debt diverges slightly more with higher workload. We would expect an alignment proportional to the invested migration costs and reduction of the number of entities for migration strategies that delay migrating legacy entities. Yet, debt is calculated based on the number of remaining legacy entities and on past on-read and on-release migration costs, the latter of which almost triples causing the relatively expensive debt for the lazy strategy. 
Thus, we can conclude that high workload levels the differences pertaining to the selected migration strategies as regards both migration costs and latency. % and slightly amplifies the migration debt. 

\begin{observation}
In terms of the variance of workload intensity, the \emph{eager} strategy represents an upper bound with respect to cumulated migration costs and remains invariant as a lower bound with respect to latency, and vice versa for the \emph{lazy} strategy in terms of upper/lower bounds. % and migration debt. 
It also holds that with higher workload in between releases compared to lower workload:
\begin{compactitem}
\item The cumulated migration costs diverge less with more queries being served, thus less costs can be saved especially in case of the \emph{lazy} and \emph{predictive} strategies, and (very) slightly less with the \emph{incremental} strategy. %The efficiency of migration cost investment for the \emph{predictive} strategy depends on the workload as well as on the prediction set size.
\item Latency also diverges slightly less, i.e., improves for migration strategies that delay migrating legacy entities.
%\item Migration debt diverges slightly more.
\item The opposite is true for lower compared to higher workload.
\end{compactitem}
\end{observation}

\subsection{Schema Modification Operations (SMOs)}
\label{subsec:complexity}

In the course of schema evolution, different kinds of schema changes occur. They can be distinguished by how many classes, tables, or entity types the data model changes affect~\cite{Curino2008,Scherzinger2013}. \emph{Single-type} SMOs affect exactly one entity type, specifically, these are \texttt{add}, \texttt{delete}, and \texttt{rename}, all of which are implemented in our middleware \Darwin{}. \emph{Multi-type} SMOs affect exactly two entity types at once, the most common are \texttt{copy}, \texttt{move}, \texttt{split}, or \texttt{merge}. The first two are investigated in our experiments. For the sake of brevity, we disregard \texttt{split} and \texttt{merge} as they can be mapped onto the implemented SMOs. The higher the percentage of multi-type operations among SMOs becomes compared to single-type operations, the more types and thus, the more entities are affected by the data model changes. We refer to this as \emph{higher multi-type SMO complexity}. 

Our experiments show that the higher the multi-type SMO complexity is, the higher the cumulated migration costs become across all migration strategies. Their increase corresponds with the amount of legacy entities that are being migrated and the number of affected types. Latency also increases for migration strategies that delay migrating entities affected by the schema changes, in our experiments the \emph{lazy}, \emph{incremental}, and \emph{predictive} strategies. This is because queries that refer to types that are affected by a schema change have to be rewritten. A higher multi-type SMO complexity, where entity types are restructured by copying and moving attributes, is oftentimes the case during major schema changes in agile development settings.

\begin{figure}[t]
\centering
\includegraphics[trim=3 50 50 5,clip,width=0.475\textwidth]{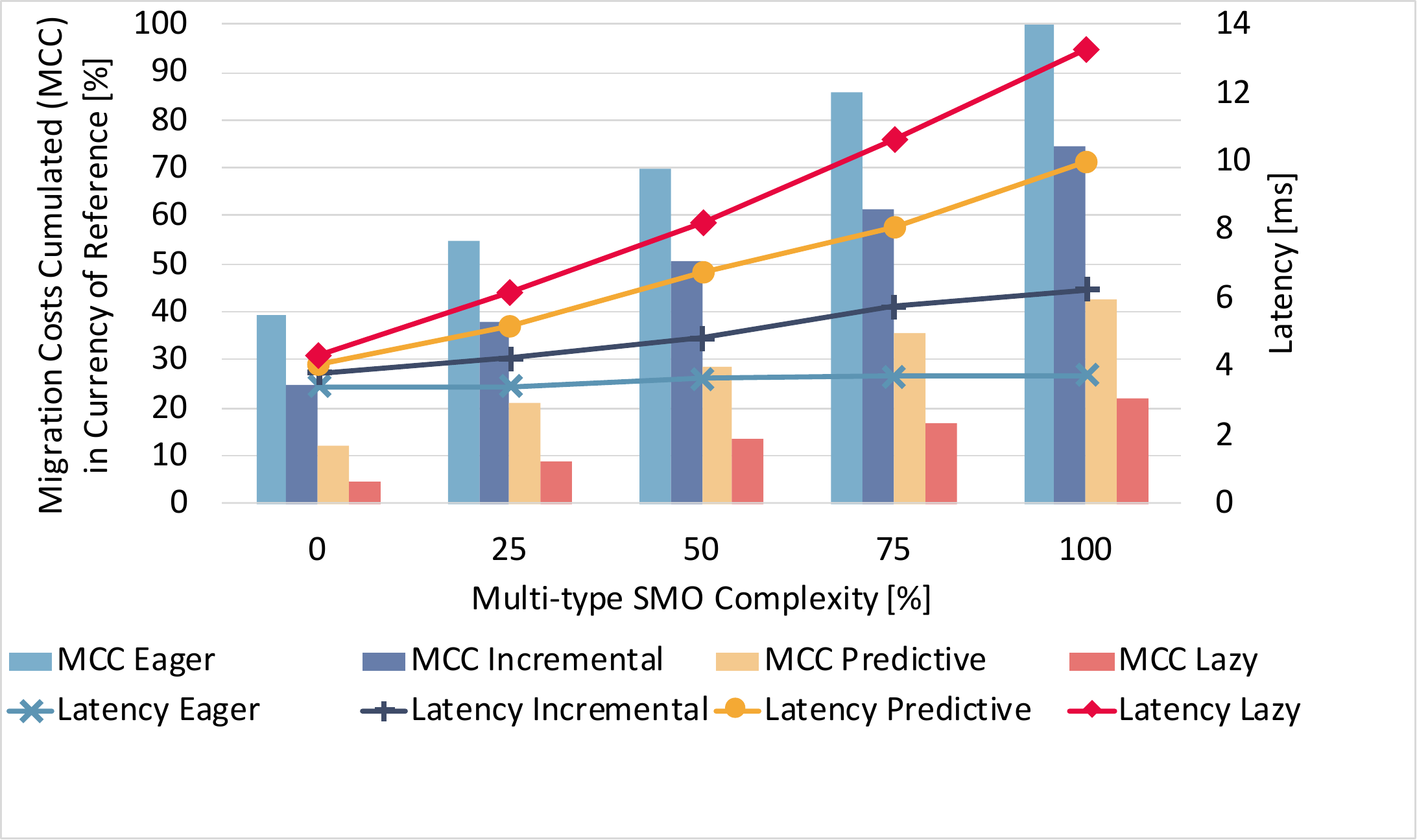}
\caption{Migration costs and latency increase proportionally with higher multi-type SMO complexity, while latency of the eager strategy remains equal at maximal migration costs; Configuration: Pareto-distributed medium workload, standard cardinality.}
\label{fig:complexity0-75MCC}
\end{figure}

In Figure~\ref{fig:complexity0-75MCC}, multi-type SMO complexity is increased in steps of 25 percentage points, i.e., 0\%, 25\%, 50\%, 75\%, 100\%, and the cumulated migration costs for the \emph{eager} strategy also increase quite proportionally by roughly $40\%$ per 25 percentage points compared to 100\% single-type operations. Thus, there is an overhead of 15 percentage points connected with multi-type SMOs, which is part of our further research. It can be noted that the increase of migration costs corresponds almost exactly to the number of affected types (factors of 1.4, 1.79, 2.19, and 2.56, i.e., with slightly declining gradient, which can be attributed to the circumstance that migration becomes slightly more efficient when entities are migrated in bulks). At the same time, the latency of the \emph{eager} strategy remains approximately equal as all entities are up-to-date. In case of the \emph{lazy} strategy it can be observed that with higher multi-type SMO complexity, the cumulated migration costs increase (1.88, 2.77, 3.54, 4.54) as well as the latency (1.42, 1.9, 2.45, 3.07, i.e., with slightly increasing gradient). Similarly, this proportionality can be noted for the \emph{incremental} and \emph{predictive} strategies as well. The cumulated migration costs of the \emph{predictive} and the \emph{incremental} strategies are located regularly in between the \emph{lazy} strategy and the \emph{eager} strategy corresponding to the increases in multi-type SMO complexity. The latency consistently reflects the investments of migration costs for both the \emph{predictive} and the \emph{incremental} strategies.

Summing up, with increasing \emph{multi-type SMO complexity}, the cumulated migration costs of all migration strategies increase almost proportionally. Although, the absolute difference between \emph{lazy} and \emph{eager} more than doubles (+128\%), the relative differences decrease from factor 8.1 in case of 0\% multi-type SMO complexity (100\% single-type SMOs) to factor 4.7 in case of 100\% (see  Figure~\ref{fig:complexity0-75MCC}). In terms of latency, the latency of the \emph{eager} strategy remains equal whereas  for the other strategies it increases proportionally to the number of affected types and invested migration costs. %In terms of migration debt the graphs of the different migration strategies diverge considerably more as with the lazy strategy the debt more than doubles (108\%) comparing 100\% single-type complexity to 100\% multi-type SMO complexity as the debt directly corresponds to the number of legacy entities, which doubles when multi-type operations affect two entity types. 
Thus, we can conclude that high multi-type SMO complexity can be considered a cost driver regarding both metrics as it greatly amplifies the absolute differences pertaining to the selected migration strategies in terms of migration costs and latency. % and migration debt.

\begin{observation}
In terms of multi-type SMO complexity of schema modification operations, the \emph{eager} strategy represents an upper bound with respect to cumulated migration costs and remains invariant as a lower bound with respect to latency, and vice versa for the \emph{lazy} strategy in terms of upper/lower bounds. % and migration debt.
It also holds with increasing multi-type SMO complexity for all migration strategies that:
\begin{compactitem}
\item The cumulated migration costs increase and diverge more with higher multi-type SMO complexity in absolute values, yet diverge less in relative values. 
More costs can be saved with migration strategies delaying migrating legacy entities when the multi-type SMO complexity is higher, although in absolute values the costs for these strategies become higher as well.
\item Latency diverges considerably more with higher multi-type SMO complexity, i.e., worsens considerably with strategies delaying migrating legacy entities proportionally to the number of affected types and invested migration costs.
%\item Migration debt diverges considerably more, i.e., directly corresponds to the number of affected types and thus legacy entities.
\item The opposite holds for decreasing multi-type SMO complexity.
\end{compactitem}
\end{observation}

\subsection{Data Model Cardinality of Relationships}
\label{subsec:cardinality}

\begin{figure}[ht]
\centering
\includegraphics[trim=11 38 6 6,clip,width=0.475\textwidth]{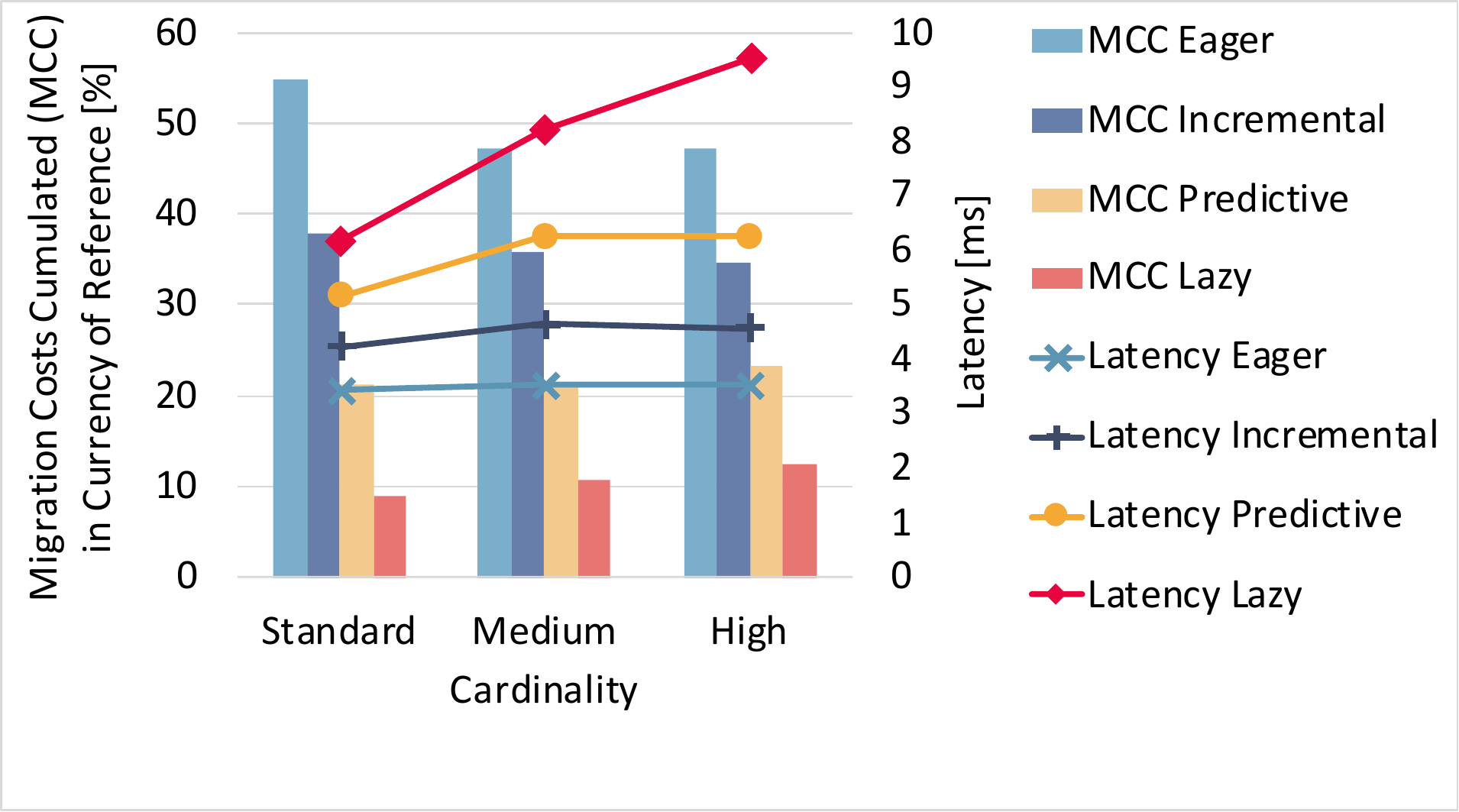}
\caption{With higher cardinality the cumulated migration costs diverge more and latency less; Configuration: Pareto-distributed medium workload, 25\% multi-type SMO complexity.}
\label{fig:card}
\end{figure}

In order to reach clarification how the data model of the persisted entities influences the resulting migration costs and latency, we have run \MCMC{} with varying cardinality of the relationships. Figure~\ref{fig:card} shows the migration costs for increasing cardinality of 1:1-, 1:10-, and 1:25-relationships, simulating standard, medium, and high cardinality of 1:n-relationships in a data model, in this subsection abbreviated as \emph{cardinality}.\footnote{For instance, in 2018 Amazon Prime members in the US placed 24 orders on average per year whereas regular customers placed 13 orders~\cite{DigitalCommerce}. This motivated our medium and high cardinality parameter instantiations. Admittedly, there are higher relationships of the cardinality, e.g., followers or postings on social media, yet, our choice is balanced as regards the simulated number of entities and the studied effects are transferable.}

In the data model with 1:1-relationships, we assume the entities to be evenly distributed over the types, whereas at higher cardinality there exist many more entities on the n-sides of the 1:n-relationships. In our experiments, the 1,000 entities are distributed over the types according to the relationships. For instance, with 1:1-relationships there are 33.4\% \emph{Player} entities, 33.3\% \emph{Mission} entities, and 33.3\% \emph{Place} entities, and with 1:25-relationships there are 0.2\% \emph{Player} entities, 4\% \emph{Mission} entities, and 96\% \emph{Place} entities. Not quite intuitively, although the number of affected entities per type changes drastically with higher cardinality, in case of single-type SMOs the overall average number of affected entities does not change as the operations affect the types equally. This holds true regardless of whether each type is filled to the maximum n specified in the relationships or not, as there is always a sampling size of 1,000 entities. Yet, if there are much less \emph{Place} entities as specified in n as there could possibly be, then the effects that we want to investigate could be mitigated. Thus, we ensured that with a number of 1,000 entities the distribution among the types comes close to the maximum n with all investigated cardinalities.

However, counter-intuitively, in case of multi-type operations higher cardinality of the relationships mean less affected entities on average by a schema change. In our experiments with 1,000 entities, the average number of entities that are affected by a multi-type SMO is approx. 667 for a 1:1-, 545 for a 1:10-, and 525 for a 1:25-relationship. In case of 25\% multi-type SMO complexity, as is the default parameter instantiation, for a 1:1-relationship approx. 416 entities are affected, for 1:10 386, and for 1:25 381, which corresponds to relative values of 93\% for 1:10 and 92\% for 1:25 compared to the reference value of a 1:1-relationship. The relative values of affected entities are calculated for 1:10- and 1:25-relationships and for each investigated percentage of multi-type SMO complexity (values in parentheses in columns 3 and 4 in Table~\ref{table:cardinality}). The measured relative migration costs are summarized in Table~\ref{table:cardinality} for the \emph{eager} strategy. For instance, the 50\% row is read as follows: Compared to the costs of 0\% multi-type SMO complexity, costs of 50\% multi-type SMO complexity are 79\% higher for 1:1-relationships, 37\% higher for 1:10 (58\% higher in theory), and 28\% higher for 1:25 (54\% higher in theory). We assume that the real measures are lower than the theoretical values due to the efficient implementation of the multi-type SMOs in connection with higher cardinalities of the relationships.

\begin{table}[t]
\begin{center}
\begin{footnotesize}
\begin{tabularx}{0.47\textwidth}{rrrr} \toprule
\emph{Multi-type SMO} & 1:1-Relationships  & 1:10-Relationships & 1:25-Relationships\\
\emph{Complexity} & Rel. Value & Rel. Value & Rel. Value \\\midrule
0\% & 100\% (Reference) & 100\%  & 100\% \\
25\% & 140\%  & 121\% (130\%)  & 121\% (129\%) \\
50\% & 179\%  & 137\% (158\%)  & 128\% (154\%) \\
75\% & 219\%  & 157\% (184\%)  & 140\% (180\%) \\
100\% & 256\% & 163\% (210\%)  & 150\% (202\%) \\
\bottomrule
\end{tabularx}
\end{footnotesize}
\end{center}
\caption{Measured migration costs w.r.t. multi-type SMO complexity and cardinality (vs. theoretic calculation).}
\label{table:cardinality}
\end{table}

Getting back to the investigation how the data model of the persisted entities impacts the resulting metrics: The cumulated migration costs have been measured in our experiments at 86\% for medium and for high cardinality (121\% with reference to 0\% multi-type SMO complexity) for the \emph{eager} strategy, which is quite close to the theoretic 93\% and 92\% (130\% and 129\%). 
As can be seen in Figure~\ref{fig:card}, the migration costs for the \emph{predictive} and the \emph{lazy} strategy tend to increase slightly for higher cardinality. Although with higher cardinality the workload concentrates on \emph{Place} entities as these are prevalent in number and thus accesses are more probable, the SMOs affect the types evenly by a third in case of single-type operations and in a 1:2:1-relationship for multi-type operations (\emph{Player}:\emph{Mission}:\emph{Place}). However, the \emph{Player} entities, of which there are very few with high cardinality, are accessed overproportionally, since we consider this to be more realistic. As a consequence, the migration costs slightly increase with strategies that delay migrating legacy entities.

\begin{figure}[ht]
\centering
\includegraphics[trim=5 7 8 4,clip,width=0.475\textwidth]{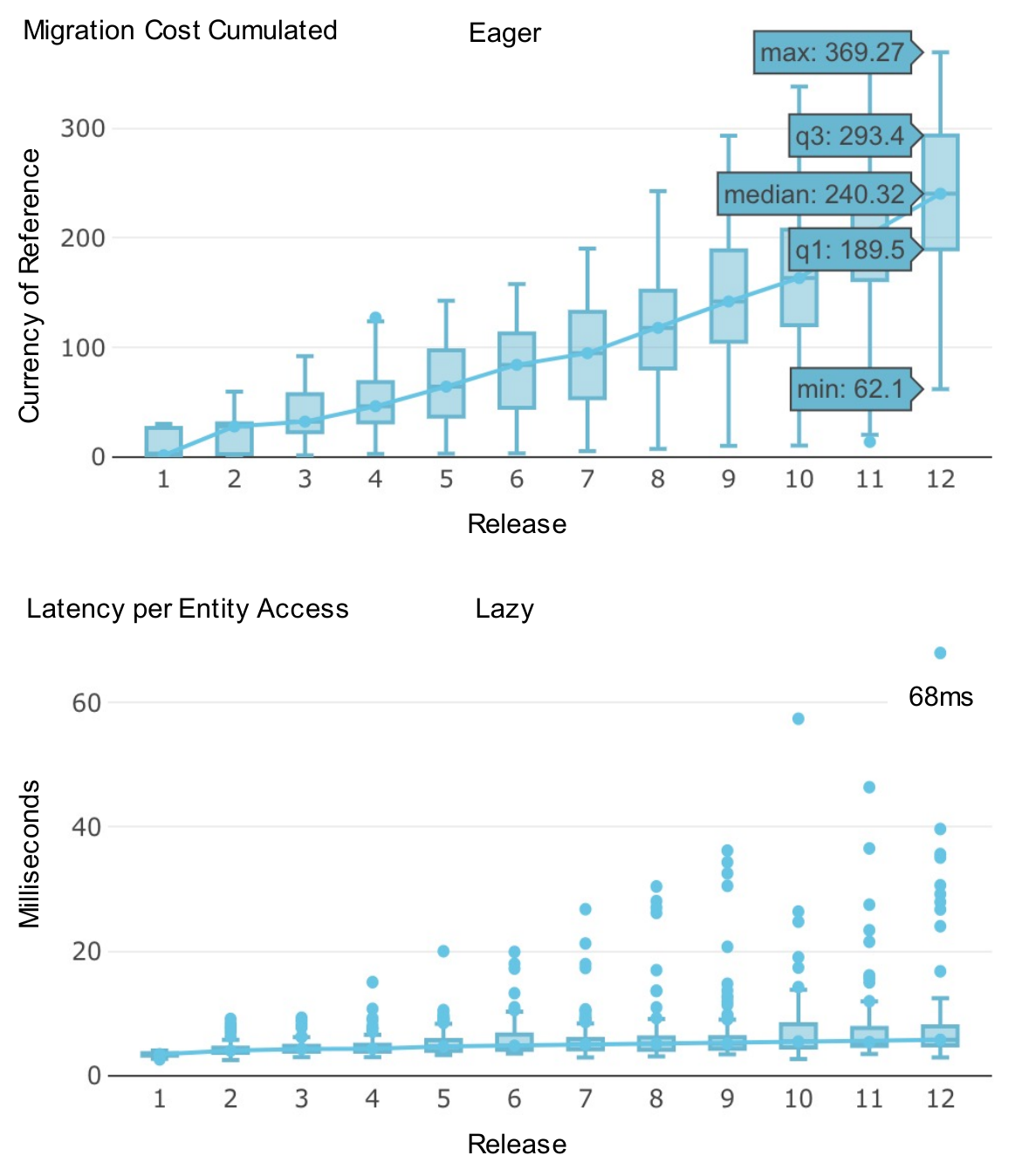}
\caption{With higher cardinality the variance of the cumulated migration costs increases particularly with the \emph{eager} strategy and latency with the \emph{lazy} strategy; Configuration same as in Figure~\ref{fig:card}.}
\label{fig:cardinality2}
\end{figure}

Due to the high variance of the metrics in case of higher cardinalities, individual results vary considerably (refer back to Figure~\ref{fig:regression}). For instance, with high cardinality, the cumulated migration costs for the \emph{eager} strategy after 12 releases are 600\% as much for the highest measured value compared to the lowest measured value (compare with upper chart of Figure~\ref{fig:cardinality2}, other strategies vary similarly). Thus, a higher safety margin should be assumed to accommodate for the variance of schema evolution in response to the higher cardinality. In case that a 50\% confidence interval is viewed as acceptable in order to comply with cost-related SLAs, migrations costs for the \emph{eager} strategy should be expected being within the boxes. Latency for the \emph{eager} strategy is then minimal and very predictable due to its low variance. %Of course, other confidence percentages can be required, too.

The averages of the measured latency in Figure~\ref{fig:card} corresponds quite consistently with the above discussions that latency responds to the invested migration costs, though in case of the \emph{lazy} strategy, it stands out that latency increases considerably with higher cardinality at mildly increasing migration costs on average. The \emph{predictive} and \emph{incremental} strategies show this effect to  lesser extent. %Regarding higher multi-type SMO complexity as detailed in Table~\ref{table:cardinality}, the cumulated migration costs for the \emph{eager} strategy came roughly as close to the theoretically calculated values. 
The example of tail latencies shows the difficulty to comply with latency-related SLAs, especially frequent with the \emph{lazy} strategy (see lower chart of Figure~\ref{fig:cardinality2}). Although the median is at 5.8ms and the 75th percentile (for a 50\% confidence interval) is at 7.9ms, we have measured tail latencies up to 68ms per one entity and with practically no external influences (note that the average means used in Figure~\ref{fig:card} respond more to outliers as explained in~\ref{sec:preliminaries}). In this case, if latency-related SLAs need to be complied with, a compromise between the metrics is potentially risky and should be considered in the decision how and when to migrate legacy entities.

Summing up, the cardinality of the relationships of the data model influences the cumulated migration costs. They are lower with higher cardinality for the \emph{eager} strategy in case that multi-type SMO complexity operations are present. For strategies utilizing the Pareto principle they slightly increase under the above discussed assumption of overproportional accesses of \emph{Player} entities. Due to a potentially high variance of the metrics, individual results vary considerably and thus, a higher safety margin should be assumed. We can conclude that higher cardinality levels the differences in terms of migration costs and amplifies them in terms of latency. 

\begin{observation}
In terms of higher cardinalities of the relationships of the data model, the \emph{eager} strategy represents an upper bound with respect to cumulated migration costs and remains invariant as a lower bound with respect to latency, and vice versa for the \emph{lazy} strategy in terms of upper/lower bounds. % and migration debt.
It also holds with higher cardinality of the relationships for all migration strategies that:
\begin{itemize}
    \item The cumulated migration costs diverge less with higher cardinality if multi-type SMO complexity operations are present. They show a high variance for all strategies.
    \item Latency diverges more and considerably increases in case of the \emph{lazy} strategy. The less a strategy migrates legacy entities, the higher becomes the variance of the latency with practically no variance for the \emph{eager} strategy.
    \item The opposite is true for lower cardinality of the relationships.
\end{itemize}
\end{observation}

%%%%%%%%%%%%%%%%%%%%%%%%%%%%%%%%%%%%%%%%%%%%%%%%%%%%%%%
%%%%%%%%%%%%%%% SECTION 6: DISCUSSION %%%%%%%%%%%%%%%%% 
%%%%%%%%%%%%%%%%%%%%%%%%%%%%%%%%%%%%%%%%%%%%%%%%%%%%%%%

\section{Discussion and Summary}
\label{sec:discussion}

We presented the results of near-exhaustive calculations of \MCMC{} by means of which software project stakeholders can remain in control of the consequences of data model evolution. We can equip them with information so that they can base their decisions during software application development on transparency regarding the impact on the metrics by selecting a migration strategy to suit the production settings. We examined the effects of the variation of migration scenario characteristics in detail, that is, distribution and intensity of workload, kinds of schema modification operations, and the cardinality of the relationships of the data model, on migration costs and latency with respect to the different migration strategies. Fig.~\ref{fig:summary} summarizes the impacts of the scenario characteristics on the metrics after 12 releases of schema changes, which we discuss in this section. Although the results are specific for typical scenarios, they are representative and constitute a heuristic that allows stakeholders to make reliable predictions in all migration scenarios.

Whether the software development is still ongoing or has shifted to its maintenance, the following development can be influenced with each further release of new software. The characteristics of the migration scenario are beyond a direct control, certainly in case of workload distribution and intensity. In case of multi-type SMO complexity and the cardinalities of the relationships a direct influence is conceivable, yet certainly not recommendable. By means of selecting a migration strategy that most likely suits the given SLAs regarding migration costs and latency, stakeholders can now make a decision that is a cost-aware and well-justified compromise. However, neither the scenario characteristics nor their impact can be predicted and thus, SLA compliance might be in jeopardy. In this case, the impact can still be mitigated by changing how often a new release is rolled out productively, because the release frequency indirectly moderates the impact. In this section, we compare and evaluate the impact of the scenario characteristics on the metrics after 12 releases of schema changes, distill a heuristics, and summarize our findings in a handy table, which supports decision-making on a suitable migration strategy and a release strategy. 

%The characteristics of the migration scenario are beyond a direct control, certainly in case of workload distribution and intensity. In case of multi-type SMO complexity and the cardinalities of the relationships a direct influence is conceivable, yet certainly not recommendable. In an application that already exists, the characteristics of the data model should be viewed as a given. By means of selecting a migration strategy that most likely suits the given SLAs regarding migration costs and latency and thus, implies certain preferences regarding the tradeoff between them, stakeholders can now make a decision that is a cost-aware compromise. However, neither the scenario characteristics can exactly be predicted nor their impact and thus, some SLAs might still not be complied with. In this case, the impact of workload intensity and multi-type SMO complexity on the metrics can still be mitigated by changing how often a new release is rolled out productively, because the frequency of releases indirectly moderates these characteristics. Whether the software development is still ongoing or has shifted to its maintenance, the further development can be influenced with each further release of new software. In this section, we compare and evaluate the impact of the scenario characteristics on the metrics after 12 releases of schema changes, distill a heuristics, and summarize our findings in a handy table, which supports decision-making on a suitable migration strategy and a release strategy. 

If the application has a Pareto distribution typical for database applications, it has been shown that a good compromise can be achieved by the \emph{predictive} migration. For giving up on a certain amount of latency, the migration costs are reduced by a considerable measure, much more so than under the uniform distribution.  Depending on the preferences regarding the tradeoff, maximally saving of costs with the \emph{lazy} migration seems as too good of a bargain, because if the data model of the application has a high cardinality, then individual costs and latency vary considerably. Depending what the consequences of a non-compliance of SLAs would entail, a confidence interval regarding the metrics can be specified in order to be on the safe side. In comparison to the \emph{incremental} strategy, the \emph{predictive} strategy achieves a better compromise between the metrics with regard to SLA compliance, because its latency is generally more stable and predictable under both workload distributions. This is especially true if the migration increments do not match with intervals of low workload.

In Figure~\ref{fig:summary}, the column \emph{WL Distribution} summarizes the migration costs and latency as factors when the workload changes from the uniform distribution to the Pareto distribution, or vice versa. The factors of each metric can be compared within a row (one strategy) or between rows (more strategies). Column headings highlighted in orange are migration costs and  latency in blue. Comparing two values within a metric, a higher factor means an increase of that metric. The quotient of two values indicates how the characteristic amplifies or levels the differences between the migration strategies, e.g., if the workload distribution changes from uniform to Pareto, then the migration costs for the \emph{lazy} strategy change from 0.4 to 0.2 (in respect of the eager strategy, i.e., approximately half). The green columns are the default parameter settings, to which all other columns are referenced. In case that factors are especially high (red values), much migration costs must be spent or tail latencies be expected, respectively. E.g., if there is a uniform distribution of data accesses, then care must be taken regarding latency, which reaches 3.3 times the amount after 12 releases of schema changes in case of the \emph{lazy} strategy compared to the \emph{eager} strategy while still spending 40\% of the migration costs compared to \emph{eager} migration. 

In comparison with the \emph{incremental} strategy, it can be argued that under both distributions the \emph{predictive} migration strategy achieves the better compromise between the metrics with regard to SLA compliance, because its latency is generally more stable and predictable (compare Figures~\ref{fig:distribution} and~\ref{fig:cardinality2}). This is especially important if the migration increments executed by the \emph{incremental} strategy do not match with recurring patterns of low workload.

\nop{
\begin{table*}[t]
\begin{center}
\begin{tabularx}{0.9\textwidth}{lllllllll} \toprule
Scenario Characteristic& \multicolumn{2}{X}{WL Distribution} & \multicolumn{2}{X}{WL Intensity} & \multicolumn{2}{X}{Multi-type SMOs} & \multicolumn{2}{X}{Cardinality} \\\cmidrule(lr){2-3}\cmidrule(lr){4-5}\cmidrule(lr){6-7}\cmidrule(lr){8-9} %%%%%%%
& \multicolumn{2}{l}{Uniform $\rightarrow$ Pareto} & \multicolumn{2}{l}{Low $\rightarrow$ High } & \multicolumn{2}{l}{0\% $\rightarrow$ 100\%} & \multicolumn{2}{l}{1:1 $\rightarrow$ 1:25} \\\cmidrule(lr){2-3}\cmidrule(lr){4-5}\cmidrule(lr){6-7}\cmidrule(lr){8-9} %%%%%%%
Migration Strategy & Cost & Latency & Cost & Latency & Cost & Latency & Cost & Latency\\\toprule %%%%%%%
Eager & 1 $\rightarrow$ 1 & 1 $\rightarrow$ 1 & 1 $\rightarrow$ 1 & 1 $\rightarrow$ 1 & 1 $\rightarrow$ 2.6 & 1 $\rightarrow$ 1.1 & 1 $\rightarrow$ 0.9 & 1  $\rightarrow$ 1 \\\midrule %%%%%%%
Incremental & 1 $\rightarrow$ 1 & 1 $\rightarrow$ 0.7 & 1 $\rightarrow$ 1.1 & 1 $\rightarrow$ 0.9 & 1 $\rightarrow$ 3 & 1  $\rightarrow$ 1.6 & 1 $\rightarrow$ 0.9 & 1 $\rightarrow$  1.1\\\midrule %%%%%%%
Predictive??? & 1 $\rightarrow$ 0.6 & 1 $\rightarrow$ 0.6 & 1 $\rightarrow$ 2.4? & 1 $\rightarrow$ 0.9 & 1 $\rightarrow$ 3.6 & 1 $\rightarrow$ 2.3 & 1 $\rightarrow$ 1.1 & 1  $\rightarrow$ 1.2 \\\midrule %%%%%%%
Lazy & 1 $\rightarrow$ 0.4 & 1 $\rightarrow$ 0.5 & 1 $\rightarrow$ 2.3 & 1 $\rightarrow$ 0.9 & 1 $\rightarrow$ 4.5 & 1 $\rightarrow$ 3.1 & 1  $\rightarrow$ 1.3 & 1  $\rightarrow$ 1.5 \\\midrule\midrule %%%%%%%
Divergence Eager-Lazy & 2.6 $\rightarrow$ 6.1 & 3.3 $\rightarrow$ 1.8 & 9.1 $\rightarrow$ 4.1 &  2 $\rightarrow$ 1.7 & 8.1 $\rightarrow$ 4.7 & 1.3 $\rightarrow$ 3.5 & 6.1 $\rightarrow$ 3.8 & 1.8 $\rightarrow$ 2.7 \\\bottomrule
\end{tabularx}
\end{center}
\caption{Impact of 12 releases of schema changes on the metrics (incremental values averaged) (IN UEBERARBEITUNG).}
\label{fig:summary}
\end{table*}
}

\begin{figure*}[ht]
\centering
\includegraphics[trim=52 450 136 55,clip,width=\textwidth]{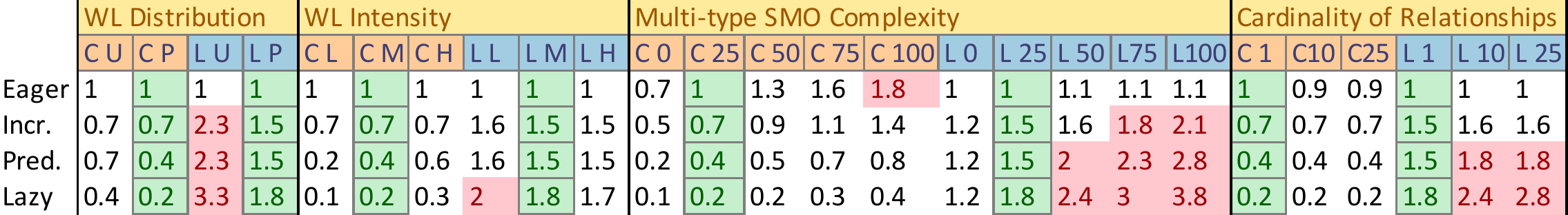}
\caption{Impact of 12 releases of schema changes on the metrics; C: Cost, L: Latency, U: Uniform, P: Pareto; columns of default settings in green and relatively high factors in red; latency for the incremental strategy is projected to the next release.}
\label{fig:summary}
\end{figure*}

As regards the distribution of the kinds of SMOs, which we refer to as \emph{multi-type SMO complexity}, the metrics respond quite proportionally to the number of affected types and in terms of the tradeoff between migration costs and latency (compare with the column \emph{Multi-type SMO Complexity} of Figure~\ref{fig:summary}). A higher multi-type SMO complexity, where entity types are restructured by copying and moving attributes, is oftentimes the case during major schema changes in agile development settings. During these phases, the stakeholders should be aware that whatever compromise has been decided for in terms of a migration strategy, migration costs and latency usually become considerably higher, with the exception of the latency with the \emph{eager} strategy which remains approximately the same, yet at much higher migration costs. In case that there are service-level agreements in place that limit the maximum amount of migration costs, then the release strategy could be adapted to plan releases in shorter intervals. This should be considered especially because the migration costs increase with increasing multi-type SMO complexity like with no other scenario characteristic (factors 1.8 for \emph{eager} through 0.4/0.2$\approx$2 for \emph{lazy} from 25\% to 100\% multi-type SMO complexity, compare with Figure~\ref{fig:summary}) and thus, high multi-type SMO complexity can be considered the most important cost driver of schema evolution. Notably, the possibility of saving costs with other than the \emph{eager} strategy should be carefully weighed with regard to tail latency that are very likely when the multi-type SMO complexity changes, especially with the \emph{lazy} strategy.

If the application has a high \emph{cardinality of relationships} of the underlying data model, then the cumulated migration costs, somewhat contrary to intuition, are on average lower for \emph{eager} and \emph{incremental} and slightly higher for \emph{lazy} and \emph{predictive} strategies compared to standard cardinality of 1:1-relationships. Yet, stakeholders should not mistake this information as a rule of thumb. Since individual costs and latency vary considerably with higher cardinalities, it should be clarified what consequences a non-compliance of service-level agreements would entail and which confidence interval should be assumed in order to be on the safe side. This is especially true for tail latency in case of the \emph{lazy} strategy and for migration costs in case of the \emph{eager} strategy, as they can increase considerably.

By multiplying the factors in the table in order to match the migration scenario, the impact can be predicted and service-level agreements fulfilled even after a migration strategy has been selected. As stated before, adapting the frequency of new software releases can then mitigate the impact of further schema changes. Of course, a longer period in between releases usually includes more schema changes, which usually entails both higher migration costs and latency, so this advice of adapting the release frequency periods has to be weighed carefully in case of longer periods in between releases.

%%%%%%%%%%%%%%%%%%%%%%%%%%%%%%%%%%%%%%%%%%%%%%%%%%%%%%%
%%%%%%%%%%%%%%% SECTION 7: CONCLUSION %%%%%%%%%%%%%%%%% 
%%%%%%%%%%%%%%%%%%%%%%%%%%%%%%%%%%%%%%%%%%%%%%%%%%%%%%%

\section{Conclusion and Outlook}
\label{sec:conclusion}

We have presented the results of our in-depth investigation how data model evolution impacts migration costs and latency while taking relevant characteristics of migration scenarios into account and identifying cost drivers. We discussed the implications of these results for software project stakeholders enabling them to remain in control of the operative costs for data model evolution and base their decision-making during software application development on transparency regarding the impact on the competing metrics migration costs and latency. We shed light on the possibilities to adapt the pace of software releases according to our findings in order to ascertain the compliance with SLAs. Furthermore, we characterized the performance of the most popular migration strategies with respect to all relevant the migration scenarios. Last but not least, we are the first to apply a probabilistic Monte Carlo method of repeated sampling which brought the complexity of data model evolution under control. 

We argued that software project stakeholders can and should base their decision-making during software application development on transparency of the impact on the competing metrics migration costs and latency with regard to the decision alternatives. With this transparency, compromises can be reached in terms of opportunity costs due to the tradeoff between migration costs and latency in accordance with service-level agreements and under consideration of the migration scenario, thus constituting a heuristic approach. Specifically, our near-exhaustive experiments clearly confirm that workload distribution has a high impact on the performance of migration strategies that delay migrating legacy entities. We discussed how workload intensity and multi-type SMO complexity can be mitigated by the release strategy. Furthermore, we have identified multi-type SMO complexity as a cost driver and despite the average impact of the data structure's cardinality being rather negligible, the high variance of potential impacts necessitates precautions with regard to SLA compliance.

These results contribute significantly towards our research goal of defining a migration advisor for all relevant migration scenarios. Further research is focusing now on applying the gained insights in order to develop a self-adaptive migration strategy which realizes the heuristics on the level of a migration strategy. This self-adaptive migration strategy could be parameterized according to the preferences on the tradeoff between migration costs and latency and their opportunity costs, based on a theoretical analysis already published in~\cite{Hillenbrand2020}. This appears especially interesting when self-adaptive strategies would be implemented in cloud solutions, which would compete with on-premise solutions in terms of a better cost efficiency, energy reduction, and sustainability. %Extending the underlying schema evolution cost model by different cloud provider pricing models is readily feasible with our middleware \Darwin{} and advisor tool \MigCast{} and can be another migration scenario to be investigated. 

\begin{acks}
This work has been funded by the Deutsche Forschungsgemeinschaft (DFG, German Research Foundation) - grant no. 385808805. We also thank Tobias Kreiter and Maksym Levchenko for their contributions to \Darwin{} and \MigCast{}.
\end{acks}

%\clearpage
\balance
\bibliographystyle{ACM-Reference-Format}
\bibliography{bib}

 \end{document}